\documentclass[3p,times,twocolumn]{elsarticle}

\usepackage{ecrc}
\usepackage{pdfpages}
\renewcommand{\theequation}{\arabic{section}.\arabic{equation}}

\def\be{\begin{equation}}
\def\ee{\end{equation}}
\def\bqa{\begin{eqnarray}}
\def\eqa{\end{eqnarray}}
\def\lsim{\raise0.3ex\hbox{$<$\kern-0.75em\raise-1.1ex\hbox{$\sim$}}}
\def\gsim{\raise0.3ex\hbox{$>$\kern-0.75em\raise-1.1ex\hbox{$\sim$}}}

\volume{00}

\firstpage{1}

\journalname{Nuclear Physics B Proceedings Supplement}

\runauth{Dieter Schildknecht}

\jid{nuphbp}

\jnltitlelogo{Nuclear Physics B Proceedings Supplement}

\usepackage{amssymb}

\usepackage[figuresright]{rotating}

\begin{document}

\begin{frontmatter}




\dochead{}

\title{The color dipole picture of low-x DIS\tnoteref{label*}
\tnotetext[label*]{Presented at
Ringberg Workshop on New Trends in HERA Physics, September 25-28,2011,
(to be published in Nucl. Phys. B, Proc. Suppl.)}}


\author{Dieter Schildknecht}

\address{Fakult\"at f\"ur Physik, Universit\"at Bielefeld,
  Universit\"atsstra\ss e 25, 33615 Bielefeld, Germany\\
and\\
Max-Planck-Institute for Physics, F\"ohringer Ring 6, 80805 Munich, Germany\\
E-Mail: Dieter.Schildknecht@physik.uni-bielefeld.de}

\begin{abstract}
Deep inelastic electron scattering (DIS) from nucleons at low values of the
Bjorken variable $x \cong Q^2/W^2 \lsim ~ 0.1$ proceeds via fluctuations
of the photon into quark-antiquark dipole states that subsequently interact 
with the gluon field in the nucleon. Dependent on the interaction energy,
$W$, the color-gauge-invariant dipole interaction with the gluon field in
the nucleon, for any fixed dipole size, contains the limits of i) color
transparency and ii) saturation, where ``saturation'' stands for the approach
to a hadronlike dipole-proton interaction cross section. All essential
features of the experimental results on low-x DIS, as a consequence of the 
color-gauge-invariant dipole interaction follow model independently i.e.
without specific ansatz for the dipole cross section. The model-independent
results in particular include the low-x scaling behavior of the photoabsorption
cross section, $\sigma_{\gamma^*p} (W^2,Q^2) = \sigma_{\gamma^*p}
(\eta (W^2,Q^2))$, with definite functional dependence on the low-x  scaling
variable $\eta (W^2,Q^2) \cong Q^2/\Lambda^2_{sat} (W^2)$ in the limits of
$\eta (W^2,Q^2) \gg 1$ and $\eta (W^2,Q^2) \ll 1$, respectively.
Consistency with the pQCD-improved parton model implies the definite
value of $C_2 \cong 0.29$ for the exponent in the ``saturation scale'',
$\Lambda^2_{sat} (W^2) \approx (W^2)^{C_2}$. The longitudinal-to-transverse
ratio of the photoabsorption cross section at large $Q^2$ has the definite
value of $R = 1/2 \rho$ with $\rho = 4/3$. For $W^2 \to \infty$ at any
fixed $Q^2$, the photoabsorption cross section converges towards a 
$Q^2$-independent saturation limit that coincides with the cross section
for $Q^2 = 0$ photoproduction. In terms of the underlying gluon distribution,
the transition from the region of validity of the pQCD-improved parton model
at $\eta (W^2,Q^2) > 1$ to the saturation region of $\eta (W^2,Q^2) < 1$
corresponds to a transition from (approximate) proportionality of the
proton structure function to the gluon distribution to a logarithmic
dependence on the gluon distribution function. 
Our specific ansatz for the dipole cross section that interpolates
between the limits of $\eta (W^2,Q^2) \gg 1$ and $\eta (W^2,Q^2) \ll 1$
describes the experimental data for the proton structure function in the
full range of $0.036~GeV^2 \le Q^2 \le 316 GeV^2$.
\end{abstract}

\begin{keyword}
Deep inelastic scattering \sep Dipole picture \sep low-x physics \sep
saturation.


\end{keyword}

\end{frontmatter}



\section{Introduction}
\label{}

As a starting point, I go back to the 60's of the last century. Photon-hadron
interactions at that time were experimentally known only within a very
restricted domain of the kinematic variables, the energy $W$ and the virtuality 
of the photon, $Q^2$. The theory was dominated by the vector-meson-dominance
picture, see ref. \cite{A} for a recent review. The real or virtual photon was
conjectured to virtually dissociate, to ``fluctuate'' in modern jargon, into
the vector mesons, $\rho^0, \omega$ and $\phi$. The subsequent interaction of
$\rho^0, \omega$, and $\phi$ with a nucleon implied ``hadronlike behavior'' of
photon-hadron interactions, well verified experimentally in e.g. vector meson 
production, the total photon-nucleon cross section and other reactions. The
generalization from the scattering on nucleons to the scattering of photons on
complex nuclei of high mass number implied hadronlike ``shadowing'' \cite{B}
as a consequence of the destructive interference of a one-step and a two-step 
reaction mechanism within the nucleus. Shadowing in the interaction of real
$(Q^2 = 0)$ photons with complex nuclei was subsequently confirmed by
experiments at DESY and SLAC.

The picture of photon-hadron interactions changed dramatically, when electron
beams of much higher energy became available at the Standford Linear
Accelerator Center. The 1969 results of the SLAC-MIT collaboration on ``deep
inelastic scattering (DIS)'' showed evidence for the scaling behavior that had
been predicted by Bjorken, and they gave rise to Feynman's interpretation of
DIS in terms of the parton model.

An alternative interpretation of the DIS results from the SLAC-MIT
collaboration, in particular in the diffraction region of low values of the
Bjorken variable -- of relevance in the present context as a starting point of
the modern point of view of the color dipole picture (CDP) -- was given by 
Sakurai and myself in 1972 \cite{Sakurai}.

\begin{figure}[h]
\begin{center}
\includegraphics[scale=.4]{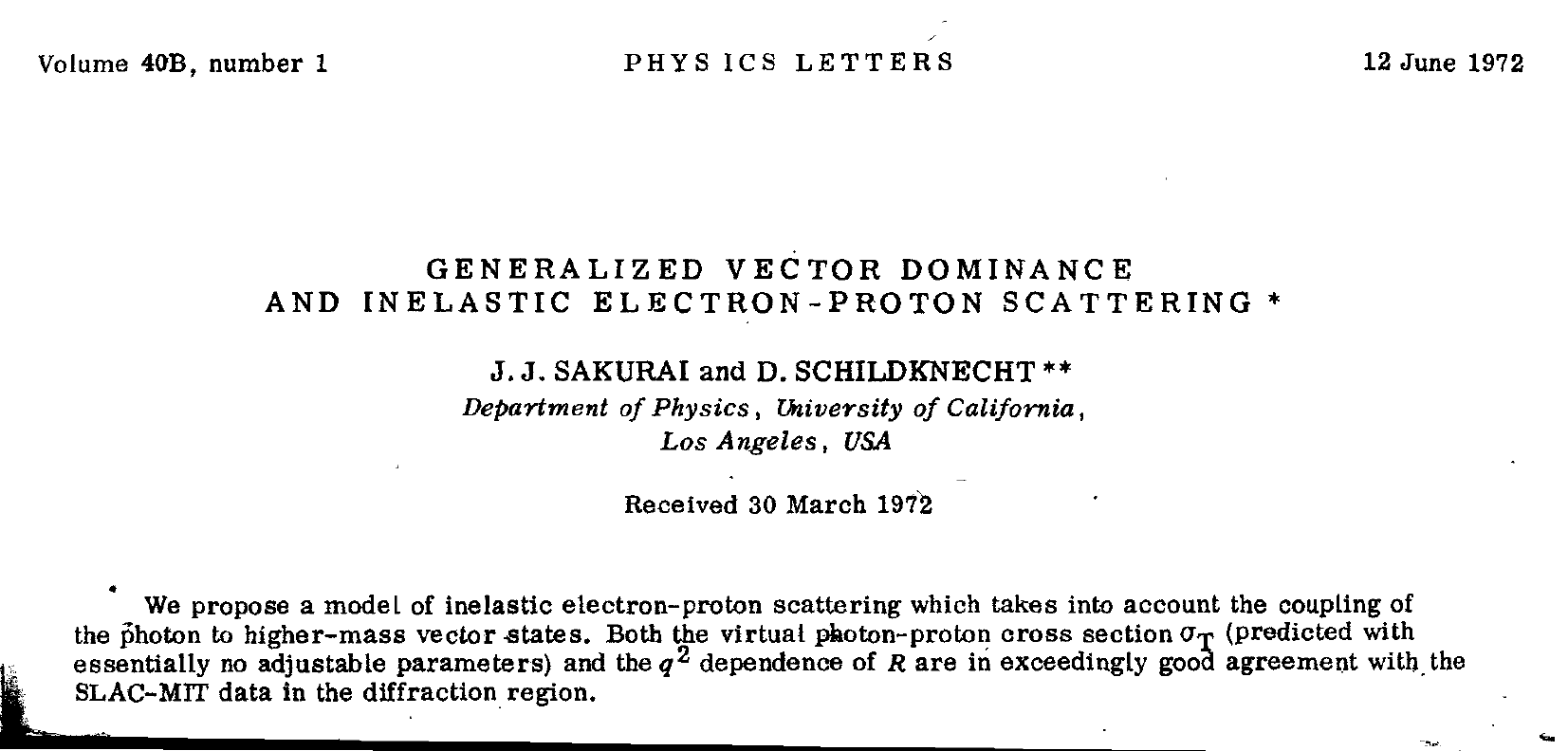}
\vspace*{-0.5cm}
\end{center}
\end{figure}

In the generalized vector dominance (GVD) approach \cite{Sakurai,Fraas}
of 1972, it was conjectured that the slow decrease with increasing $Q^2$ of the 
photoabsorption cross section observed by the SLAC-MIT collaboration was due to
so far unobserved couplings of the photon to a continuum of hadron states, more
massive than the vector mesons, $\rho^0, \omega, \phi$. Compare Fig. 1 for the
comparison of the successful GVD prediction with the DIS data from the SLAC-MIT
collaboration. 
\begin{figure}[h!]
\begin{center}
\vspace*{-0.2cm}
\includegraphics[scale=.4]{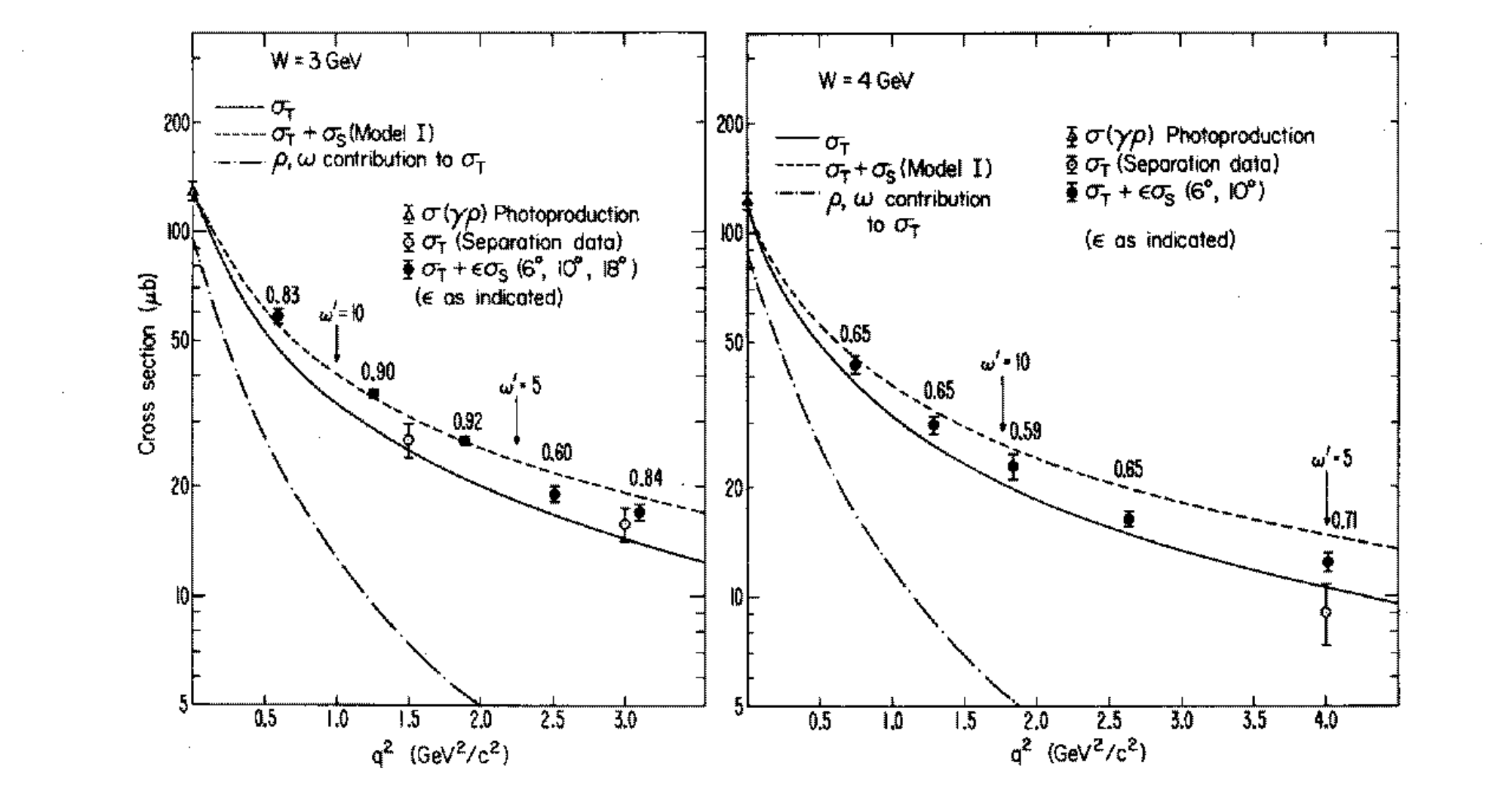}
\caption{The SLAC-MIT experimental data 
compared with GVD predictions \cite{Sakurai}}
\vspace*{-0.8cm}
\end{center}
\end{figure}

The interpretation of the DIS data in terms of photon couplings to a high-mass
continuum, $\gamma^* \to ~{\rm continuum~states}$, in addition to the $\gamma^*
\to \rho^0, \omega, \phi$ coupling required the persistence of shadowing, when
real photons were replaced by virtual ones. In the 70's and 80's of last
century, it was frequently argued that the GVD approach was invalid due to the
lack of shadowing in the scattering of virtual photons on complex nuclei. After
many years of confusion, in 1989 however, shadowing in the reaction of virtual
photons with nuclei was discovered by the EMC collaboration \cite{C}. The
results are consistent with the theoretical prediction \cite{D} from GVD,
compare Fig. 2.

\begin{figure}[h!]
\begin{center}
\includegraphics[scale=.4]{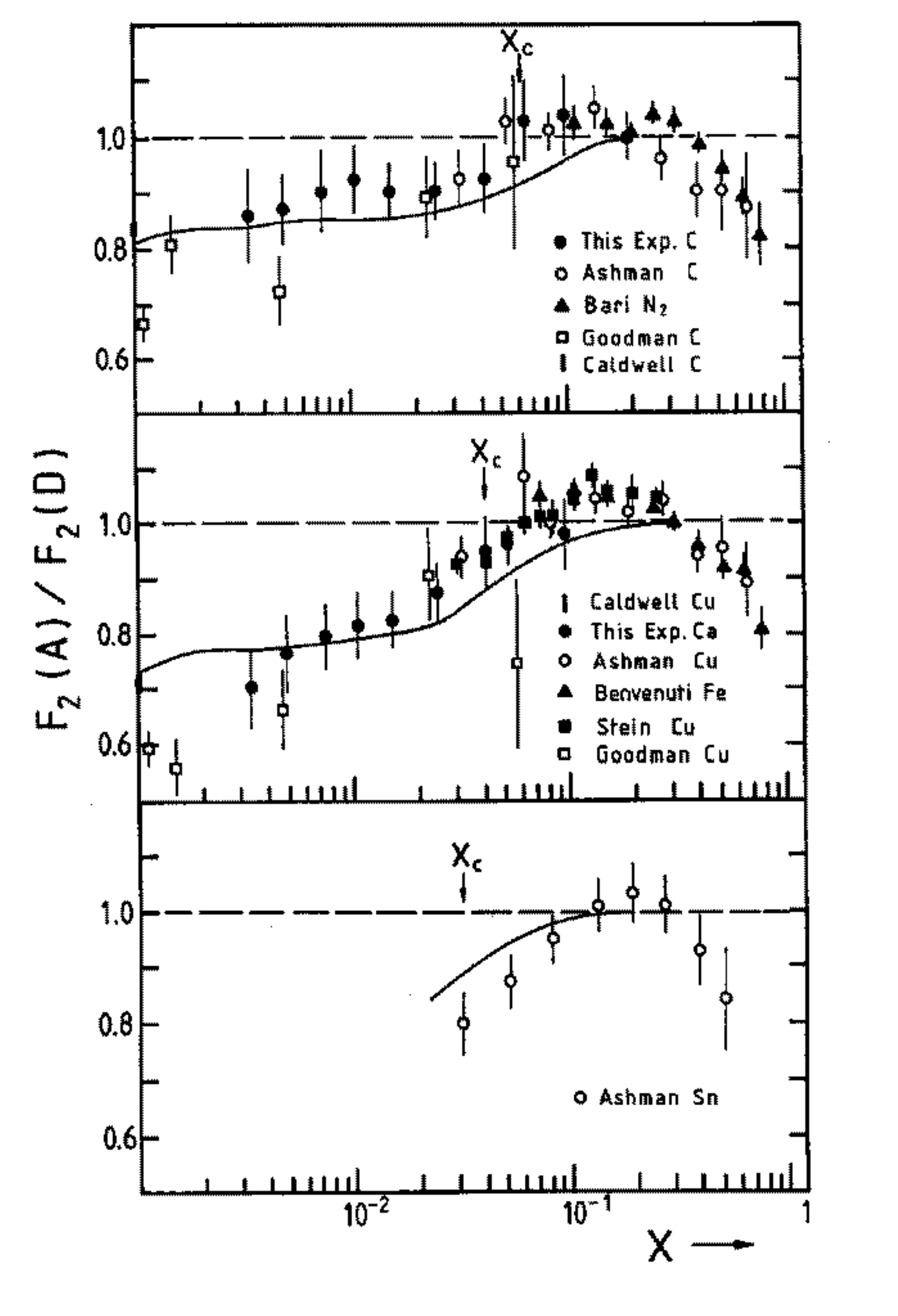}
\caption{Shadowing in the scattering of virtual photons from complex nuclei
compared with GVD predictions \cite{D}}
\end{center}
\end{figure}

Shadowing, being due to destructive interference between a one-step and a
two-step process, for virtual photons, $Q^2 > 0$, requires the existence of
diffractive production of states more massive than $\rho^0, \omega,
\phi$. Accordingly, it came without surprise that immediately after the start
of HERA, in 1994, the existence of high-mass diffractive production
(``large-rapidity-gap events'') was established at HERA.

\section{The modern picture of DIS at low x: the color dipole picture}
\renewcommand{\theequation}{\arabic{section}.\arabic{equation}}
\setcounter{equation}{0}

As in GVD, in the modern approach, the photoabsorption
reaction proceeds in two steps, i) the
$\gamma^* \to q \bar q$ transition and ii) the $(q \bar q)$-proton interaction,
whereby taking into account
\begin{itemize}
\item[i)] the internal structure of the $q \bar q$ system via the variable $0
  \le z \le 1$ that specifies the longitudinal momentum distribution between
  the quark and the antiquark the photon fluctuates into, compare Fig. 3, and

\begin{figure}[h]
\begin{center}
\includegraphics[scale=.5]{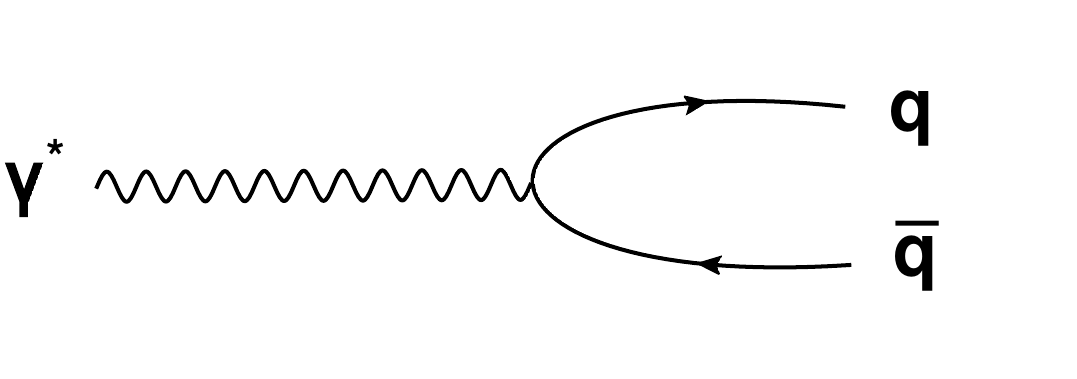}
\vspace*{-0.5cm}
\caption{The $\gamma^* q \bar q$ transition}
\end{center}
\end{figure}

\item[ii)] the $q \bar q$ interaction with the gluon field in the nucleon 
 \cite{Low}  as a
  gauge-invariant color-dipole interaction, compare the
  (virtual) forward Compton scattering amplitude in Fig.\ 4.
\end{itemize}

\begin{figure}[h]
\begin{center}
\includegraphics[scale=.5]{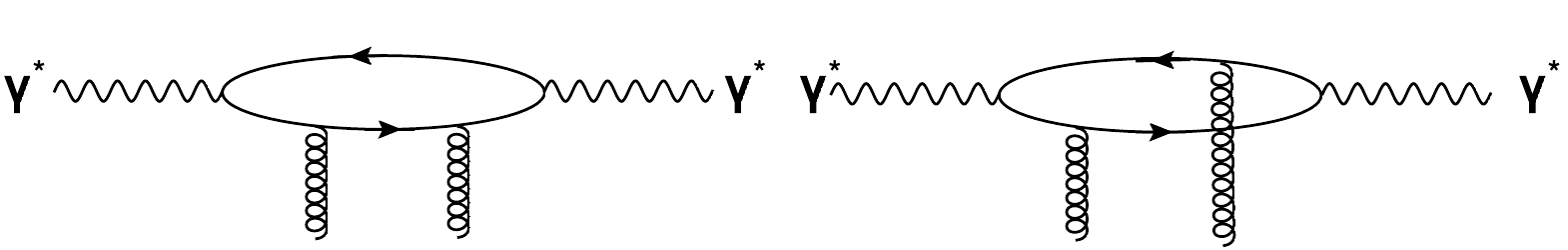}
\caption{Two of the four diagrams for the $q \bar q$ dipole interaction with
the gluon field in the nucleon}
\end{center}
\end{figure}

After Fourier transform to transverse position space, the photoabsorption 
cross section at low $x \simeq Q^2/W^2 < 0.1$ takes the form \cite{Nikolaev,
Cvetic}
\bqa
&\hspace*{-1cm}\sigma_{\gamma^*_{L,T}} (W^2, Q^2)  = \int dz \int d^2 \vec r_\bot
\vert \psi_{L,T} (\vec r_\bot, z (1 - z), Q^2) \vert^2 \cdot
\nonumber \\
&\hspace*{-1cm} \cdot \sigma_{(q\bar q)p}
(\vec r_\bot, z (1 - z), W^2). 
\label{2.1}
\eqa
The quantity $\vert \psi_{L,T} (\vec r_\bot, z (1-z),Q^2) \vert^2$ may be
interpreted as the probability for a longitudinally or a transversely polarized
photon, $\gamma^*_{L,T}$, of virtuality $Q^2$ to undergo a transition to a
$q \bar q$ state, $\gamma^*_{L,T} \to q \bar q$, being characterized by the
transverse size $\vec r_\bot$ and by the distribution of the longitudinal
momenta of quark and antiquark determined by $z(1-z)$. In the rest frame of
the $q \bar q$ fluctuation of mass $M_{q \bar q}$, the quantity $z(1-z)$ 
determines \cite{Cvetic} the direction of the quark (antiquark) with respect 
to the photon direction. The interaction cross section of the 
$q \bar q$ dipole state in
(2.1) is denoted by $\sigma_{(q \bar q)p} (\vec r_\bot, z(1-z), W^2)$. It 
depends on the energy, $W$, \cite{Cvetic, DIFF2000, Forshaw, Ewerz} of the
$(q \bar q)p$ interaction, since the photon fluctuates into an on-shell
$q \bar q$ state of mass $M_{q \bar q}$ that subsequently interacts with
the nucleon. For generality, a dependence on $z (1-z)$ is allowed for in
the dipole-proton cross section.

The gauge-invariant two-gluon interaction ii) of the $q \bar q$ dipole enters
the photoabsorption cross section in (2.1) via \cite{Nikolaev, Cvetic}
\bqa
&&\hspace*{-1cm} \sigma_{(q \bar q)p} (\vec r_\bot, z(1-z), W^2) = \nonumber \\
&&\hspace*{-1cm} = \int d^2 \vec l_\bot \tilde{\sigma}
(\vec l^{~2}_\bot, z(1-z), W^2) 
\left(1-e^{-i~ \vec l_\bot \cdot \vec r_\bot}\right).
\label{2.2}
\eqa
In (\ref{2.2}), $\vec l_\bot$ stands for the transverse momentum of the
absorbed gluon, and the first and the second term in the parenthesis on the
right-hand side in (\ref{2.2}), respectively, corresponds to the first and the
second diagram in\break Fig. 2.

For the subsequent discussions, it will be useful to equivalently 
rewrite \cite{MKDS, E} the
photoabsorption cross section (\ref{2.1}), in terms of dipole states, 
$(q \bar q)^{J=1}_{L,T}$, that describe longitudinally and 
transversely polarized $q \bar q$ states of fixed spin $J=1$ and polarization
index $L$ and $T$. In terms of the corresponding dipole cross section, 
$\sigma_{(q \bar q)^{J=1}_{L,T}p} (\vec r^{~\prime}_\bot, W^2)$, where
$\vec r^{~\prime}_\bot = \vec r_\bot z (1-z)$, the photoabsorption cross
section
(\ref{2.1}) becomes \cite{MKDS, E} 
\bqa
&&\hspace*{-1cm}\sigma_{\gamma^*_{L,T}p} (W^2, Q^2) = \nonumber \\
&&\hspace*{-1cm} \frac{\alpha}{\pi} \sum_q Q^2_q Q^2 
\int dr^{\prime 2}_\bot K^2_{0,1} (r^\prime_\bot Q) 
\sigma_{(q \bar q)^{J=1}_{L,T} p} (r^{\prime 2}_\bot, W^2).
\label{2.3}
\eqa
In the transition from (\ref{2.1}) to (\ref{2.3}), assuming massless quarks, we
inserted the explicit expressions for $\vert \psi_{L,T} (\vec r_\bot,
z (1-z), Q^2)\vert^2$ in terms of the modified Bessel functions $K_0
(r^\prime_\bot Q)$ and $K_1 (r^\prime_\bot Q)$, and $Q$ stands for 
$Q \equiv \sqrt{Q^2}$.The sum over the squared charges of the actively
contributing quarks is given by $\sum_q Q^2_q$, and the 
cross section $\sigma_{(q \bar q)^{J=1}_{L,T}} (r^\prime_\bot, W^2)$ is
related to the dipole cross section in (\ref{2.1}) by an appropriate
projection.

In terms of the $J=1$ projection, $\sigma_{(q \bar q)^{J=1}_{L,T}p}
(r^\prime_\bot, W^2)$,\break  of the dipole cross section in (\ref{2.1}),
with\break $\vec l^{~\prime 2}_\bot = \vec l^{~ 2}_\bot/z (1-z)$, the
two-gluon-coupling structure of the dipole cross section in (\ref{2.2})
becomes
\bqa
&& \sigma_{(q \bar q)^{J=1}_{L,T}p} (r^\prime_\bot, W^2)  = \nonumber \\
&& \pi \int d \vec l^{~\prime 2}_\bot \bar \sigma_{(q \bar q)^{J=1}_{L,T} p}
(\vec l^{~\prime 2}_\bot , W^2) \cdot \nonumber \\
&& \cdot \left( 1 - \frac{\int d \vec l^{~\prime 2}_\bot 
\bar \sigma_{(q \bar q)^{J=1}_{L,T} p} (\vec l^{~\prime 2}_\bot, W^2) J_0
(l^\prime_\bot r^\prime_\bot)}{\int d \vec l^{~\prime 2}_\bot
\bar \sigma_{(q \bar q)^{J=1}_{L,T} p} (\vec l^{~\prime 2}_\bot, W^2)}
\right), 
\label{2.4}
\eqa
where $J_0 (l^\prime_\bot r^\prime_\bot)$ denotes the Bessel function with
\break index $0$.

Two distinct $W$-dependent limits of the dipole cross section (\ref{2.4}) will
be relevant \cite{E} and important for the ensuing discussions. We assume
that the integrals in (\ref{2.4}) exist and are determined by a
$W$-dependent restricted range of $\vec l^{~\prime 2}_\bot < \vec
l^{~\prime_2}_{Max} (W^2)$, in which $\bar \sigma_{(q \vec q)^{J=1}_{L,T} p} 
(\vec l^{~\prime 2}_\bot, W^2)$ is appreciably different from zero, 
$\vec l^{~\prime 2}_{\bot~ Max} (W^2)$ increasing with increasing $W^2$.
The resulting cross section for a dipole of fixed transverse size, 
$r^\prime_\bot$,
strongly depends on the variation of the phase $l^\prime_\bot r^\prime_\bot$:
\begin{itemize}
\item[A)] For
\be
0 < l^\prime_\bot r^\prime_\bot < l^\prime_{\bot~Max} (W^2) r^\prime_\bot
\ll 1,
\label{2.5}
\ee
upon employing the expansion
\be
J_0 (l^\prime_\bot r^\prime_\bot) \cong 1 - \frac{1}{4} (l^\prime_\bot
r^\prime_\bot)^2 + \frac{1}{4^3} (l^\prime_\bot r^\prime_\bot)^4 + \cdots ,
\label{2.6}
\ee
we find strong destructive interference between the two additive contributions
to the $J=1$ dipole cross section (\ref{2.4}) which correspond to the first and
the second diagram in Fig. 4. The dipole cross section (\ref{2.4}) becomes
proportional to $r^{\prime 2}_\bot$ (``color transparency'' 
limit \cite{Nikolaev})
\bqa
& & \hspace*{-1cm}\sigma_{(q \bar q)^{J=1}_{L,T} p} (r^{\prime 2}_\bot, W^2) =
\label{2.7} \\
& &\hspace*{-1cm}= \frac{1}{4} r^{\prime 2}_\bot \sigma_L^{(\infty)} (W^2)
\Lambda^2_{sat} (W^2) \cdot
\left\{ \matrix{ 1,\cr
\rho_W ,} \right. \nonumber \\  
&&\left(r^{\prime 2}_\bot \ll
\frac{1}{l^{\prime 2}_{\bot~Max} (W^2)}\right),
\nonumber
\eqa
where by definition the $W^2$-dependent scale $\Lambda^2_{sat} (W^2)$ reads
\bqa
&& \Lambda^2_{sat} (W^2) \equiv \frac{1}{\sigma^{(\infty)}_L(W^2)} \pi \cdot  
\nonumber \\
&& \cdot\int d \vec l^{~\prime 2}_\bot
\vec l^{~\prime 2}_\bot \bar \sigma_{(q \bar q)^{J=1}_L p} 
(\vec l^{~\prime 2}_\bot , W^2),
\label{2.8}
\eqa
and $\sigma^{(\infty)}_L (W^2)$ is explicitly defined by (\ref{2.11}) below.
For vanishing size, $\vec r^{~\prime 2}_\bot$, the $q \bar q$ color dipole
(obviously) has a vanishing cross section. The expression for the factor
$\rho_W$ in (\ref{2.7}) is explicitly obtained by comparison of (\ref{2.7})
with (\ref{2.4}). According to (\ref{2.7}), and anticipating $\rho_W > 1$,
transversely polarized $(q \bar q)^{J=1}$ states interact with enhanced
transverse size \cite{Ku-Schi, E}
\be
\vec r^{~\prime 2}_\bot \rightarrow \rho_W \vec r^{~\prime 2}_\bot
\label{2.9}
\ee
relative to longitudinal ones. The ratio $\rho_W$ will be shown to be a
$W$-independent constant of definite magnitude, $\rho_W = \rho = 4/3$.
\item[B)] For the case of
\be
l^\prime_{\bot Max} (W^2) r^\prime_\bot \gg 1,
\label{2.10}
\ee
alternative to (\ref{2.5}), for any fixed value of $r^\prime_\bot$, rapid
oscillations of the Bessel function in (\ref{2.4}) lead to a vanishingly 
small contribution of the second term in (\ref{2.4}) thus implying
(``saturation'' limit)
\bqa
&&\hspace*{-1cm}\sigma_{(q \bar q)^{J=1}_{L,T} p} (r_\bot^{~\prime 2}, W^2) 
\cong \pi
\int d \vec l^{~\prime 2}_\bot \bar \sigma_{(q \bar q)^{J=1}_{L,T} p} 
(\vec l^{~\prime 2}_\bot , W^2) 
\equiv  \nonumber \\
&&\hspace*{-1cm} \equiv \sigma^{(\infty)}_{L,T} (W^2),~~~~
\left(r^{\prime 2}_\bot \gg
\frac{1}{l^{\prime 2}_{\bot~Max} (W^2)} \right).
\label{2.11}
\eqa
The high-energy limit in (\ref{2.11}) of sufficiently large $W$ at fixed
dipole size $r^\prime_\bot$, according to (\ref{2.4}), coincides with the
limit of  sufficiently large $r^\prime_\bot$ at fixed $W$ i.e.
\be
\lim_{r^{\prime 2}_\bot \to \infty \atop W = const} 
\sigma_{(q \bar q)^{J=1}_{L,T}p} (r^{\prime 2}_\bot, W^2) = 
\sigma_{L,T}^{(\infty)} (W^2).
\label{2.12}
\ee
At fixed dipole size, with sufficiently large energy, we arrive at the
large-dipole-size limit of the (at most weakly $W$-dependent) hadronic
cross section $\sigma^{(\infty)}_{L,T} (W^2) \simeq \sigma^{(\infty)}$.
\end{itemize}

The photoabsorption cross section in (\ref{2.3}), due to the strong 
decrease of the modified Bessel functions $K_{0,1} (r^\prime_\bot~Q)$ with
increasing argument $r^\prime_\bot Q$, is strongly dominated and actually 
determined at any fixed value of $Q^2$ by values of $r^{\prime~2}_\bot$
such that $r^{\prime 2}_{\bot} Q^2 < 1$. Whether color transparency of the
dipole cross section according to (\ref{2.5}) and (\ref{2.7}) or, 
alternatively, saturation
according to (\ref{2.10}) and
(\ref{2.11}) is relevant for $\sigma_{\gamma^*_{L,T}p} (W^2,Q^2)$
in (\ref{2.3}) depends on whether $Q^2 \gg \Lambda^2_{sat} (W^2)$ or 
$Q^2 \ll \Lambda^2_{sat} (W^2)$ is realized at a specific value of $W$.
Upon substitution of (\ref{2.7}) and, alternatively, of (\ref{2.11}) into
(\ref{2.3}), for $\sigma_{\gamma^*p} (W^2,Q^2) = \sigma_{\gamma^*_L p}
(W^2,Q^2) + \sigma_{\gamma^*_Tp} (W^2,Q^2)$ one finds \cite{DIFF2000, E}
\bqa
&& \hspace*{-1cm}\sigma_{\gamma^*p} (W^2, Q^2)  =  \sigma_{\gamma^*p} (\eta
(W^2, Q^2)) = \frac{\alpha}{\pi} \sum_q Q^2_q  \cdot
\label{2.13} \\
&& \hspace*{-1cm} \cdot \left\{ \matrix{
\frac{1}{6} ( 1 +2 \rho ) \sigma^{(\infty)}  \frac{1}{\eta(W^2, Q^2)},
~~~ (\eta (W^2,Q^2) \gg 1), \cr
\sigma^{(\infty)}
\ln \frac{1}{\eta (W^2, Q^2)},~~~~~~~~~~~~~~(\eta (W^2, Q^2) \ll 1), } \right.
\nonumber
\eqa

where we have introduced the low-x scaling variable
\be
\eta (W^2, Q^2) = \frac{Q^2 + m^2_0}{\Lambda^2_{sat} (W^2)},
\label{2.14}
\ee
and anticipated $\rho_W = {\rm const.} = \rho$. In (\ref{2.14}), via
quark-hadron duality \cite{Sakurai, ST}, we introduced the 
lower bound $m^2_0 \lsim ~ m^2_\rho$ on the masses of the
$q \bar q$ fluctuations
$M^2_{q \bar q} \ge m^2_0$, only relevant in the limit of $Q^2 \to 0$.

From the above derivation, leading to (\ref{2.13}), it has become clear
that DIS at low x proceeds via two different reaction channels.
They correspond to the first and the second diagram in Fig. 4. For
sufficiently large $Q^2 \gg \Lambda^2_{sat} (W^2)$ both channels are
open, resulting in strong destructive interference between them. With
decreasing $Q^2$ at fixed $W$, or with increasing $W$ at fixed $Q^2$,
for $Q^2 \ll \Lambda^2_{sat} (W^2)$, the second channel becomes closed,
no destructive interference any more. Only the first channel remains open,
implying that the proportionality of the photoabsorption cross section
to $\Lambda^2_{sat} (W^2)$ turns into the (soft) energy dependence
proportional to $\ln \Lambda^2_{sat} (W^2)$, compare (\ref{2.13}).

The longitudinal-to-transverse ratio of the photoabsroption cross sections
$\sigma_{\gamma^*_L p} (W^2,Q^2)$ and $\sigma_{\gamma^*_T p} (W^2,Q^2)$ at
large $Q^2 \gg \Lambda^2_{sat} (W^2)$ according to (\ref{2.13}) is given by
\bqa
&& R(W^2,Q^2)_{Q^2 \gg \Lambda^2_{sat} (W^2)}  = \nonumber \\
&& \frac{\sigma_{\gamma^*_L} (W^2,Q^2)}{\sigma_{\gamma^*_T} (W^2,Q^2)} 
\bigg|_{Q^2 \gg \Lambda^2_{sat} (W^2)} = \frac{1}{2 \rho}
\label{2.15}
\eqa
The factor 2 in (\ref{2.15}) originates from the difference in the
photon wave functions in (\ref{2.3}). The interaction with enhanced
transverse size of $(q \bar q)^{J=1}_T$ states relative to
$(q \bar q)^{J=1}_L$ states, $\rho_W$ in (\ref{2.9}), is a consequence
of the ratio of the average transverse momenta of the quark (antiquark)
in the $(q \bar q)^{J=1}_T$ state relative to the quark (antiquark) in
the $(q \bar q)^{J=1}_L$ state. Upon applying the uncertainty principle, one
obtains \cite{Ku-Schi, E}
\be
\rho_W = \rho = \frac{4}{3}.
\label{2.16}
\ee
The longitudinal structure function, with (\ref{2.15}) and (\ref{2.16}), 
at large $Q^2$
is related to the transverse one via
\be
\hspace*{-0.7cm}F_L (x,Q^2) = \frac{1}{1+2 \rho} F_2 (x,Q^2) = 
0.27 F_2 (x,Q^2).
\label{2.17}
\ee 
The result is consistent with the experimental data, compare Fig. 5.

\begin{figure}[h]
\begin{center}
\includegraphics[scale=.3]{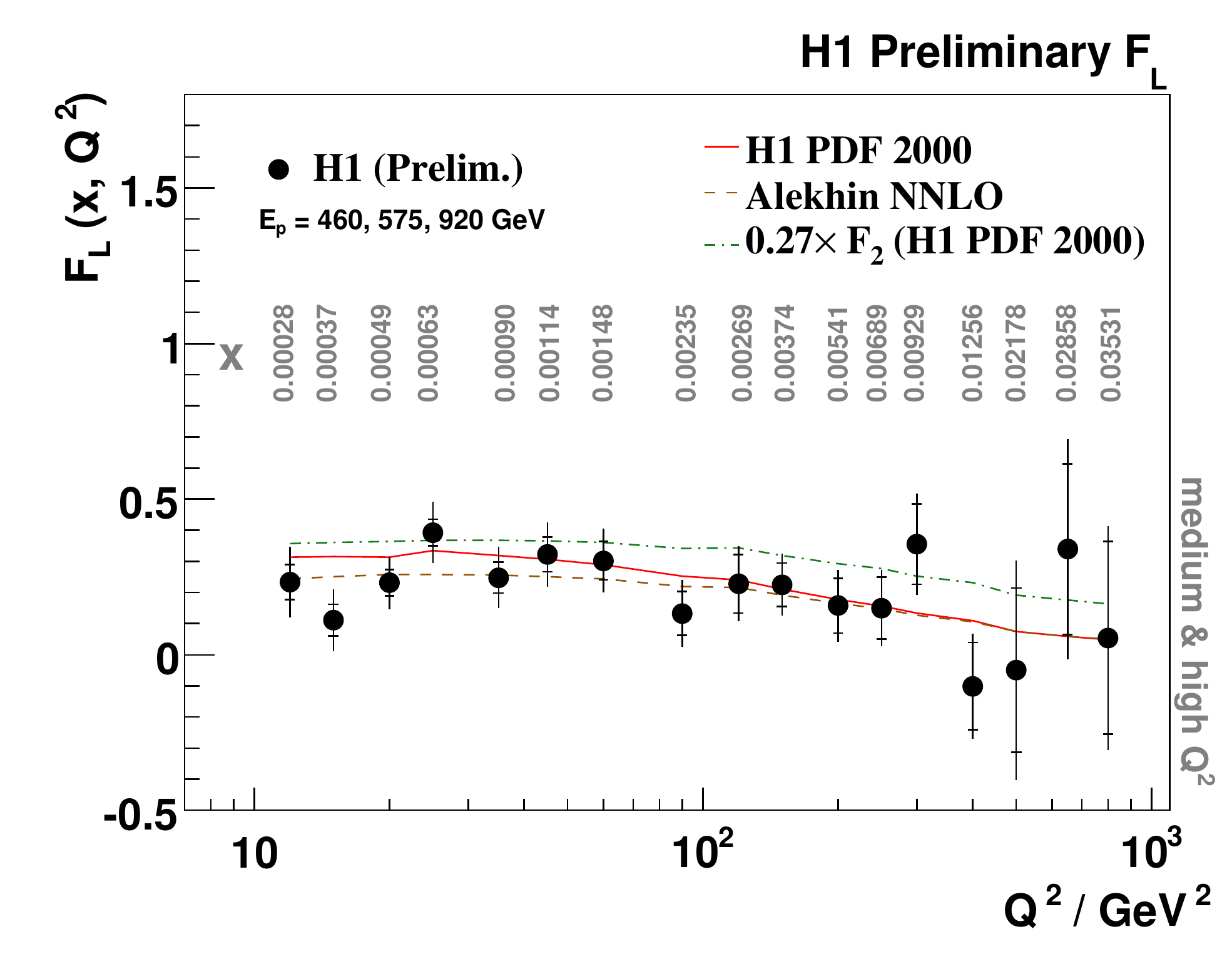}
\includegraphics[scale=.3]{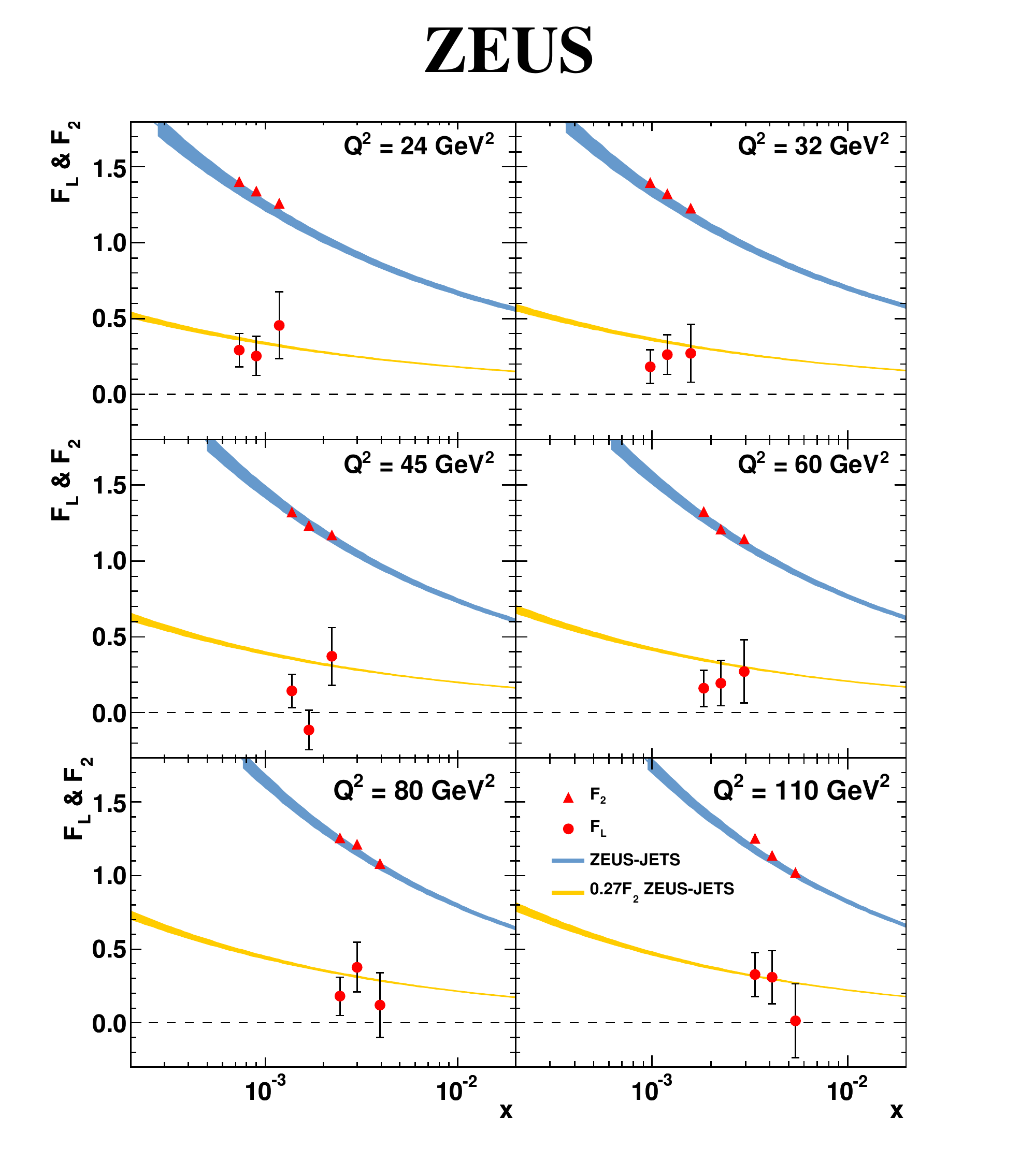}
\caption{The prediction (\ref{2.17}) compared with the experimental
data for $F_L$ and $F_2$.}
\end{center}
\end{figure}
\begin{figure}[h!]
\begin{center}
\vspace*{-0.5cm}
\includegraphics[scale=.38]{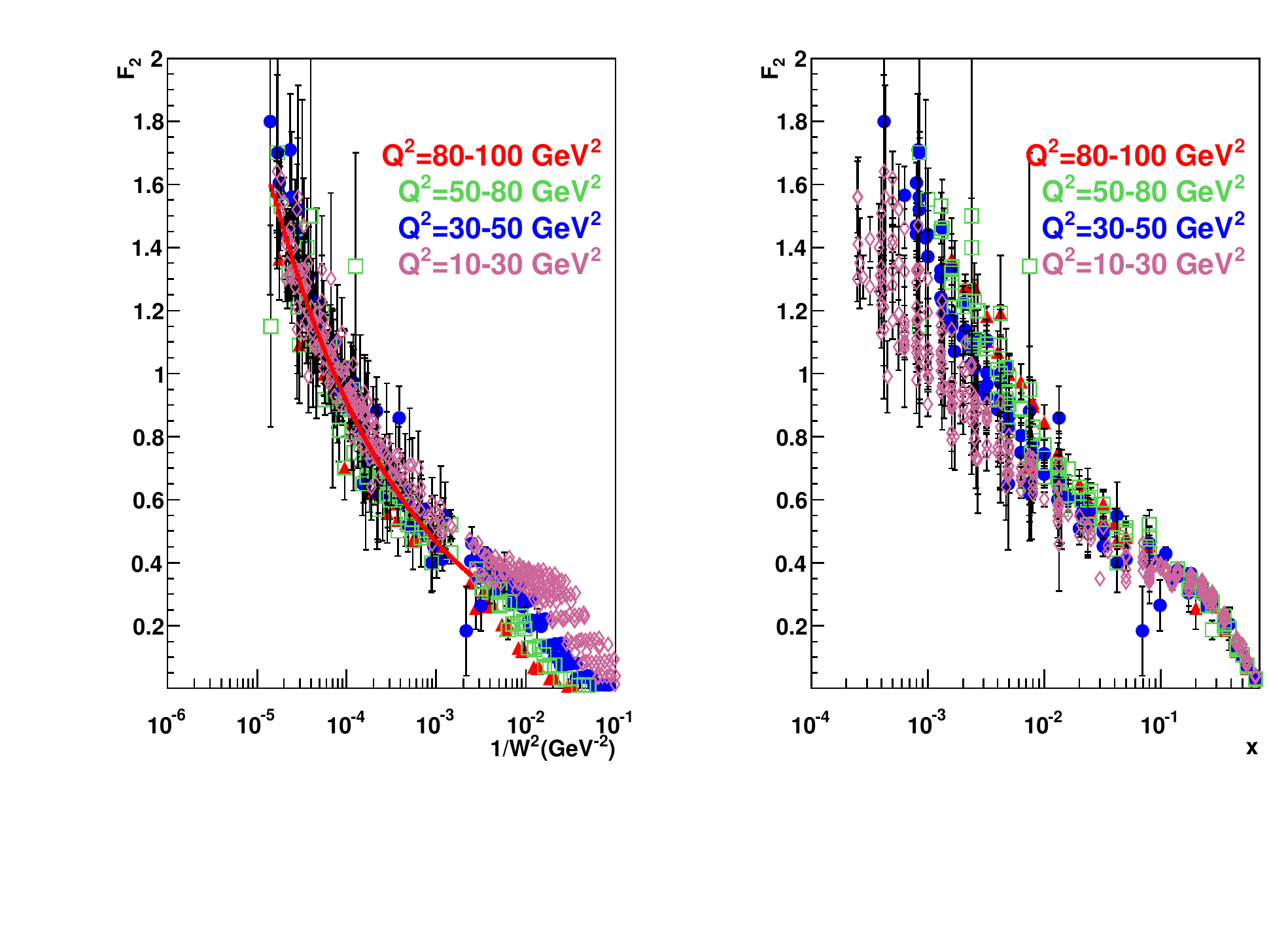}
\vspace*{-1cm}
\caption{The structure function $F_2 (x, Q^2)$ as a function of
$1/W^2$ and as a function of $x$}
\vspace*{-0.5cm}
\end{center}
\end{figure}

The $W$-dependence in (\ref{2.1}) and (\ref{2.3}) of the dipole cross
section, combined with the $1/Q^2$ dependence in (\ref{2.13}) at
large $Q^2$, implies that the structure function $F_2 (x,Q^2) \simeq
(Q^2/4 \pi^2 \alpha) (\sigma_{\gamma^*_L p} (W^2, Q^2) + 
\sigma_{\gamma^*_T p} (W^2,Q^2))$ becomes a function of the single variable
$W^2$. The experimental data in Fig. 6, in the relevant range of
$x \simeq Q^2/W^2 < 0.1$, approximately corresponding to $1/W^2 \le
10^{-3}$, indeed show the expected tendency to lie on a single line 
\cite{E} \footnote{The representation of the experimental data in
Fig. 6 was kindly prepared by Prabhdeep Kaur.}. An eye-ball fit to the
data in Fig. 6 yields
\be
F_2 (W^2) = f_2 \cdot \left( \frac{W^2}{1{\rm GeV}^2} \right)^{C_2 = 0.29}  
\label{2.18}
\ee
with $f_2 = 0.063$. We will come back to representation (\ref{2.18}) of
$F_2(W^2)$ below. For comparison, in Fig. 6, we also show $F_2(x,Q^2)$
as a function of $x$.

In Fig. 7, we show the experimental evidence for the\break low-$x$ scaling
behavior, $\sigma_{\gamma^*p}(W^2,Q^2)=\sigma_{\gamma^*p}(\eta (W^2,Q^2))$
of (\ref{2.13}), with $\eta(W^2,Q^2)$ from (\ref{2.14}), first obtained in
ref. \cite{DIFF2000} by a model-independent analysis. The experimental data
confirm this general prediction of the CDP of scaling in $\eta (W^2,Q^2)$
in the form (\ref{2.13}), which is independent of any specific ansatz for the
dipole cross section. The theoretical curve in Fig. 7 is due to the 
ansatz \cite{DIFF2000, E}
for the dipole cross section to be discussed in Section 4.

\begin{figure}[h]
\begin{center}
\includegraphics[scale=.35]{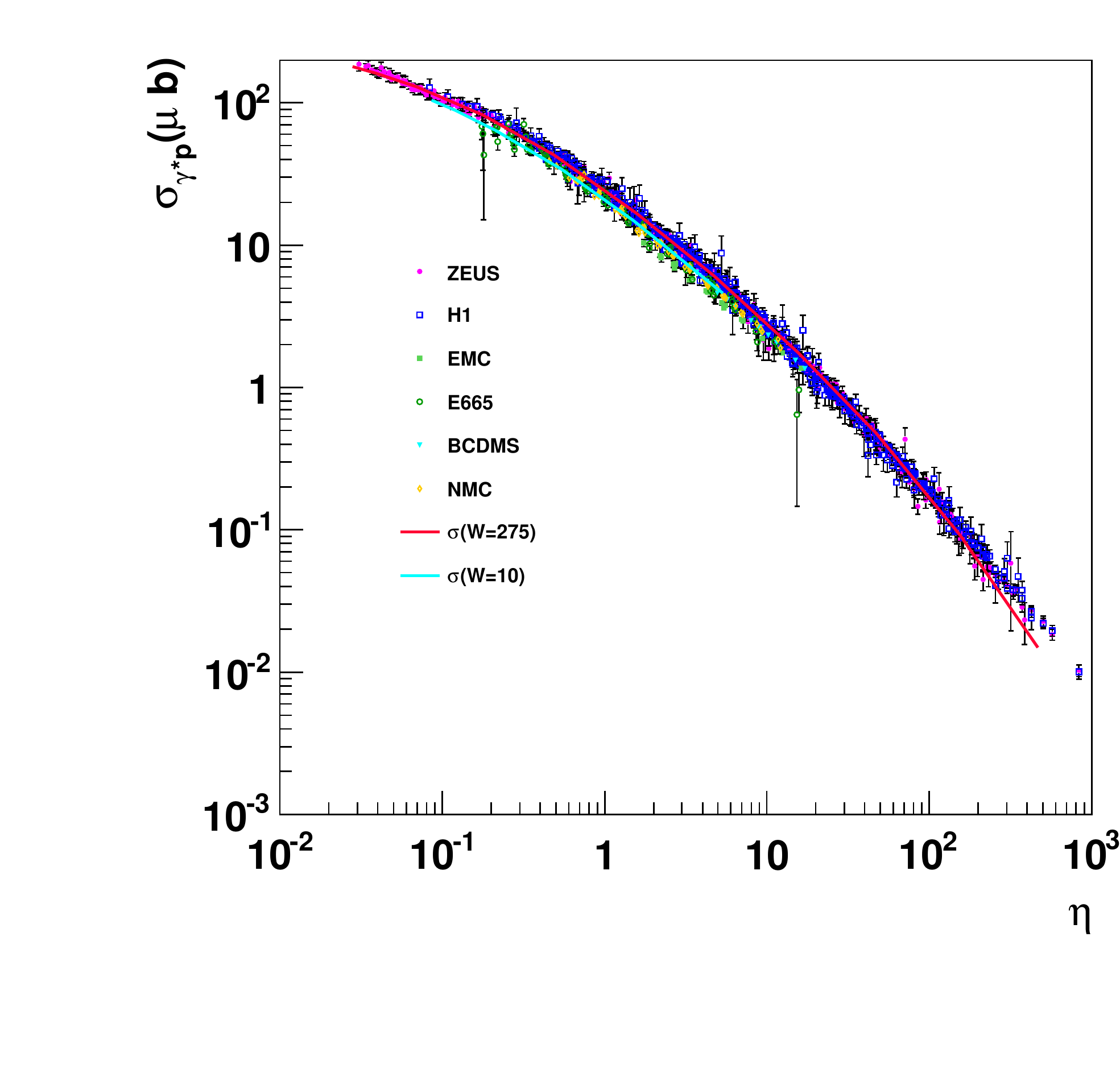}
\vspace*{-1.2cm}
\caption{Scaling of $\sigma_{\gamma^*p} (W^2,Q^2) = \sigma_{\gamma^*p} (\eta
(W^2,Q^2))$.}
\vspace*{-0.5cm}
\end{center}
\end{figure}

The logarithmic dependence of the photoabsorption cross section in
(\ref{2.13}), for $W^2 \to \infty$ at fixed $Q^2$ leads to \cite{SCHI}
\bqa
&& \lim_{W^2 \to \infty \atop {Q^2~{\rm fixed}}} 
\frac{\sigma_{\gamma^*p} (\eta (W^2, Q^2))}{\sigma_{\gamma^*p} (\eta
(W^2, Q^2 = 0))} \nonumber \\
&& = \lim_{W^2 \to \infty \atop {Q^2~{\rm fixed}}} 
\frac{\ln \left( \frac{\Lambda^2_{sat}(W^2)}{m^2_0} 
\frac{m^2_0}{(Q^2 + m^2_0)} \right)}{\ln 
\frac{\Lambda^2_{sat} (W^2)}{m^2_0}} = \nonumber \\
&& = 1 + \lim_{W^2 \to \infty \atop {Q^2~{\rm fixed}}} 
\frac{\ln \frac{m^2_0}{Q^2 + m^2_0}}{\ln 
\frac{\Lambda^2_{sat} (W^2)}{m^2_0}} = 1.
\label{2.19}
\eqa
In this limit of (\ref{2.19}), the photoabsorption cross section tends
to a $Q^2$-independent limit that coincides with $Q^2 = 0$ 
photoproduction. The convergence to this limit is extremely slow. Compare
Fig. 8, where this limit is seen in terms of the structure function
$F_2 (x,Q^2)$,
\be
\lim\limits_{{W^2\rightarrow\infty}\atop{Q^2 {\rm fixed}}}
\frac{F_2 (x\cong Q^2 / W^2, Q^2)}{\sigma_{\gamma p} (W^2)} =
\frac{Q^2}{4\pi^2\alpha} .
\label{2.20}
\ee 
The theoretical curve in Fig. 8 is due to the concrete ansatz for the dipole
cross section in Section 4 that interpolates between the regions of
$\eta (W^2,Q^2) \gg 1$ and $\eta (W^2,Q^2) \ll 1$.

\begin{figure}[h]
\begin{center}
\vspace*{-0.4cm}
\includegraphics[scale=.35]{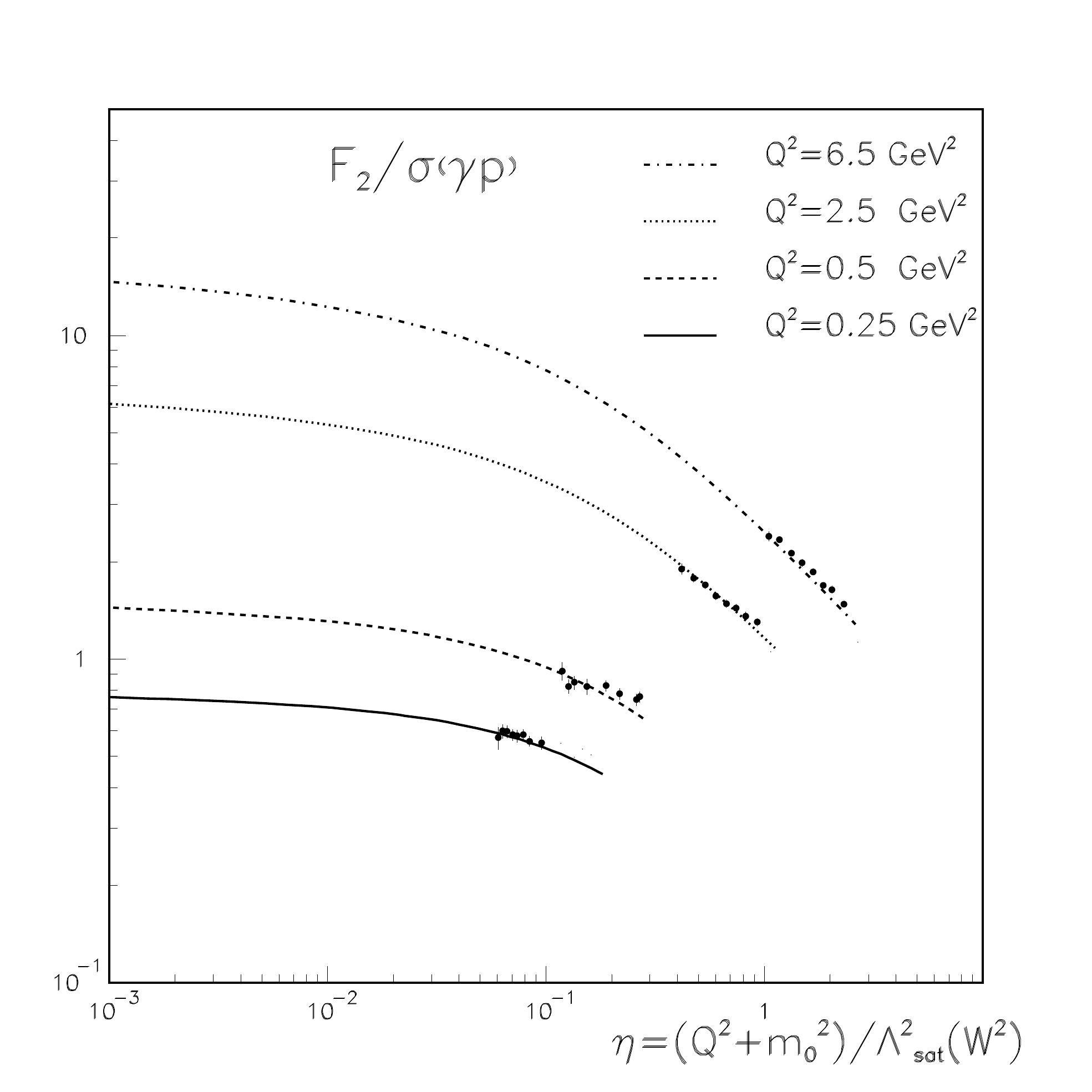}
\vspace*{-0.3cm}
\caption{The approach to saturation}
\vspace*{-0.5cm}
\end{center}
\end{figure}
An approach to a $Q^2$-independent limit for $W^2 \to \infty$ at
fixed $Q^2$ was recently independently observed by Caldwell\cite{CAL}
based on a purely empirical fit to the experimental data given by
\be
\sigma_{\gamma^*p} (W^2,Q^2) = \sigma_0 (Q^2)
\left( \frac{1}{2M_p} \frac{W^2}{Q^2} \right)^{\lambda_{eff} (Q^2)}.
\label{2.21}
\ee
The straight lines in Fig. 9 meet at a value of $W^2$ approximately
given by $W^2 \simeq 10^9 Q^2$, consistent with the above conclusion 
from the CDP.
\begin{figure}[h!]
\begin{center}
\vspace*{-3.4cm}
\includegraphics[scale=.35]{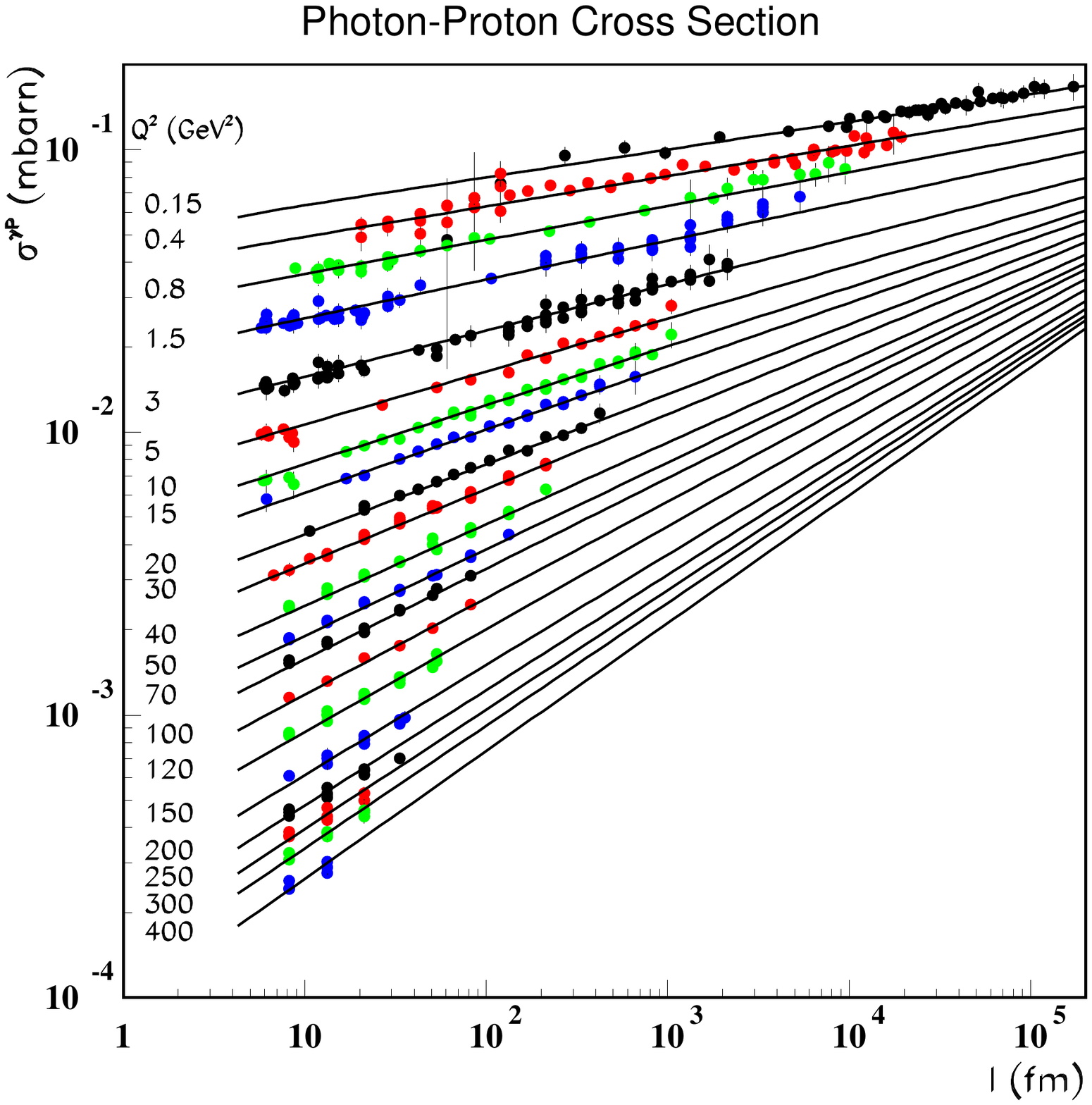}
\vspace*{-0.3cm}
\caption{The Caldwell fit (\ref{2.21})\cite{CAL}}
\end{center}
\end{figure}

The results from the above general analysis of the CDP lead to the
simple structure of the $(Q^2, W^2)$ plane shown in Fig. 10. 

\begin{figure}[h!]
\begin{center}
\vspace*{-0.5cm}
\includegraphics[scale=.35]{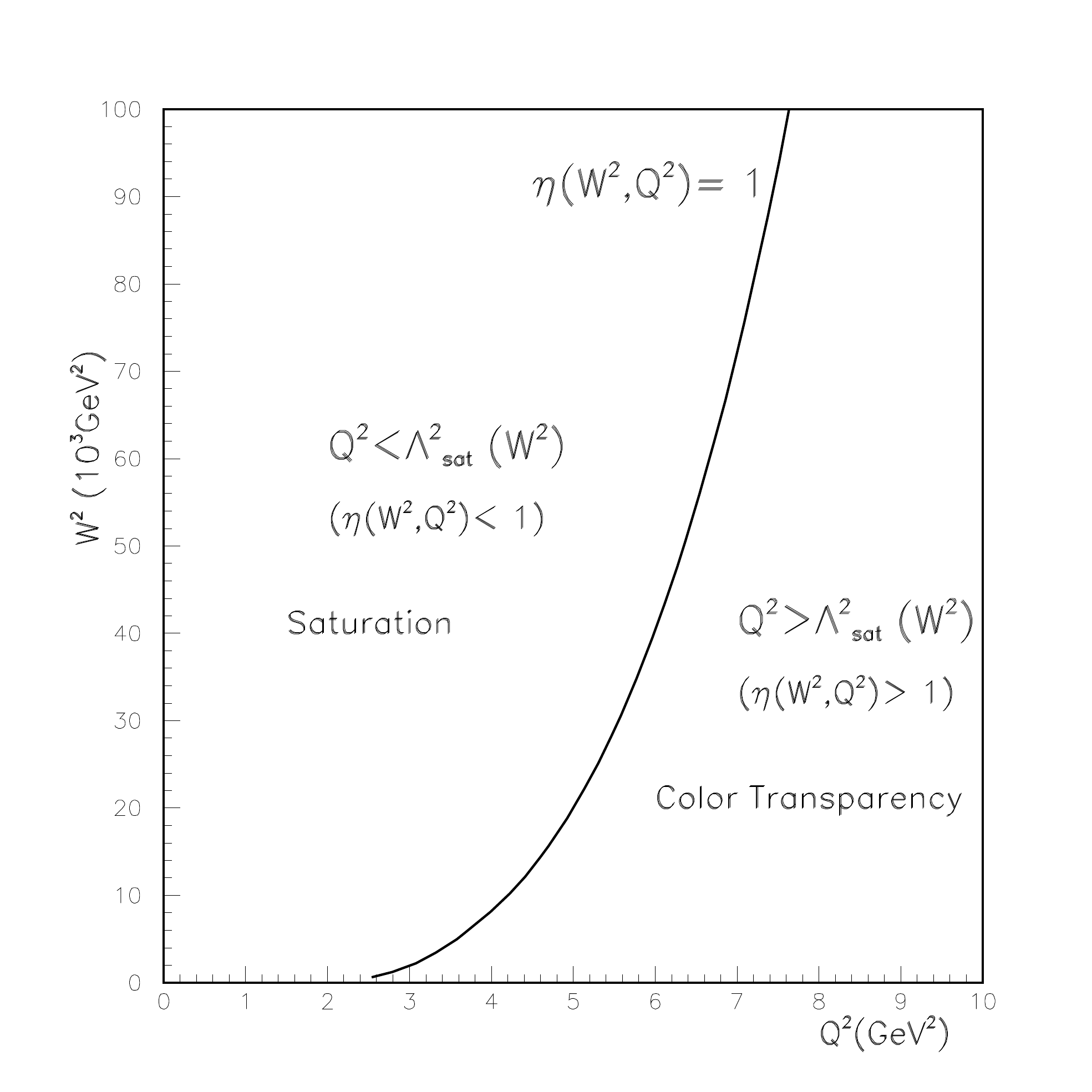}
\vspace*{-0.4cm}
\caption{The $Q^2,W^2)$ plane of the CDP.The line $\eta (W^2, Q^2) = 1$
subdivides the $(Q^2,W^2)$ plane into the saturation region of $\eta (W^2,Q^2)
< 1$ and the color transparency region of $\eta (W^2,Q^2) > 1$.}
\end{center}
\end{figure}

The $(Q^2,W^2)$ plane is subdivided into only two regions as a consequence of
the two interaction channels corresponding to the diagrams in Fig. 4.
For $\eta (W^2,Q^2) > 1$, both interaction channels, channels 1
and 2, are open, implying color transparency of the $(q \bar q)p$
dipole interaction with strong destructive interference. For
$\eta(W^2,Q^2) \ll 1$, channel 2 becomes closed. The lack of
destructive interference leads to a saturation of the cross section 
determined by the dipole interaction solely via channel 1. The
virtual-photon-nucleon cross section approaches a $Q^2$-independent
saturation limit that coincides with $Q^2 = 0$ photoproduction.

\section{The CDP, the gluon distribution function and evolution}
\renewcommand{\theequation}{\arabic{section}.\arabic{equation}}
\setcounter{equation}{0}

The CDP of DIS corresponds\footnote{With respect to this Section,
compare also ref. \cite{23a}} to the low $x$ approximation of the
pQCD-improved parton model in which the interaction of the (virtual)
photon occurs by interaction with the quark-antiquark sea in the 
proton via $\gamma^* gluon \to q \bar q$ fusion.

The longitudinal structure function in this approximation, for a wide
range of different gluon distributions, becomes proportional to the
gluon density at a rescaled value $x/\xi_L$ \cite{Martin}
\be 
F_L ( \xi_L x, Q^2) = \frac{\alpha_s (Q^2)}{3\pi} \sum_q Q^2_q G (x, Q^2),
\label{3.1}
\ee
where $G(x,Q^2) \equiv xg (x,Q^2)$ and $g(x,Q^2)$ stands for the gluon
distribution function. The rescaling factor has the preferred value of
$\xi_L \simeq 0.40$.

The structure function $F_2 (x,Q^2)$ at low $x$ in this approximation
is proportional to the sea-quark distribution, and again for a wide
range of different gluon distributions, the evolution of the structure
function $F_2 (x,Q^2)$ with $Q^2$ is determined by \cite{Lipatov, Prytz}
\be
\frac{\partial F_2 (\xi_2 x , Q^2)}{\partial \ln Q^2} = \frac{\alpha_s
  (Q^2)}{3\pi} \sum_q Q^2_q G  (x ,Q^2),
\label{3.2}
\ee
where the preferred value of $\xi_2$ is given by $\xi_2 \simeq 0.50.$

According to the CDP, compare (\ref{2.13}), and supported by the 
experimental data in Fig. 6,
the structure function $F_2 (x,Q^2)$ for sufficiently large $Q^2$ becomes
a function of the single variable $W^2$,
\be
F_2 (x, Q^2) = F_2 (W^2 = \frac{Q^2}{x}) . 
\label{3.3} 
\ee
Employing the proportionality of $F_L(x,Q^2) = (1/(1+2 \rho_W)) F_2(x,Q^2)$
from (\ref{2.7}) and (\ref{2.17}), and combining (\ref{3.1}) and 
(\ref{3.2}), upon inserting (\ref{3.3}) the evolution equation becomes
\be
(2 \rho_W + 1) \frac{\partial}{\partial \ln W^2} F_2 \left( \frac{\xi_L}{\xi_2}
  W^2 \right) = F_2 (W^2). 
\label{3.4}
 \ee
A potential dependence of $\rho_W$ on the energy $W$ is allowed in 
(\ref{3.4}).

We specify $F_2 (W^2)$ by adopting the power law \cite{E}
\be
F_2 (W^2) \sim (W^2)^{C_2} = \left( \frac{Q^2}{x} \right)^{C_2} . 
\label{3.5}
\ee
A power law in $(1/x)^\lambda$ with $\lambda$ occurs e.g. in the
``hard Pomeron solution'' \cite{Adel} of DGLAP evolution as well
as in the ``hard Pomeron'' part of Regge phenomenology with
$(1/x)^{\epsilon_0}$ and $\epsilon_0 \simeq 0.43$ from a fit
\cite{Dom}. The CDP in (\ref{3.5}) is more specific, however,
since the $W$ dependence of $F_2(W^2)$ implies that the $x$ dependence
and the $Q^2$ dependence are intimately related to each other.

Substitution of (\ref{3.5}) into (\ref{3.4}) implies the constraint
\cite{E}
\be
(2\rho_W + 1) C_2 \left( \frac{\xi_L}{\xi_2} \right)^{C_2} = 1 . 
\label{3.6}
\ee
If, and only if $\rho_W = \rho = {\rm const.}$, also the exponent
is constant, $C_2 = {\rm const.}$ With the CDP result of $\rho =
4/3$ from (\ref{2.16}), we obtain the unique value of
\be
C_2 = \frac{1}{2\rho + 1} \left( \frac{\xi_2}{\xi_L} \right)^{C_2} = 
0.29,
\label{3.7}
\ee 
where $\rho = 4/3$ and $\xi_2/\xi_L = 1.25$ was inserted. The result
for $C_2 = 0.29$ is fairly insensitive under variation of $\xi_2/\xi_L$.
For $1 \le \xi_2/\xi_L \le 1.5$, one obtains $0.27 \le C_2 \le 0.31$.
The value of $C_2 = 0.29$ is consistent with the experimental data,
compare (\ref{2.18}) and Fig. 6.

Imposing consistency between the CDP and the pQCD-improved parton
model, we thus arrived at the prediction of a definite value of
$C_2 = 0.29$ that coincides with the experimental findings in Fig. 6.

The underlying gluon distribution function can now be deduced from
(\ref{3.1}) by expressing the longitudinal structure function in terms
of $F_2 (x,Q^2)$ according to (\ref{2.17}) and inserting the power law
(\ref{2.18}),
\bqa
&&\hspace*{-1cm} \alpha_s (Q^2) G (x, Q^2)   = \nonumber \\
&&\hspace*{-1cm} =\frac{3\pi}{\sum_q Q^2_q (2\rho + 1)} 
\frac{f_2}{\xi_L^{C_2 = 0.29}} \left(
  \frac{W^2}{1 {\rm GeV}^2} \right)^{C_2 = 0.29}.~~~~~~~~~~~~~~~~~ 
\label{3.8}
\eqa
With $\rho = 4/3$ from the CDP, compare (\ref{2.16}),
the result (\ref{3.8}) contains the
single free fitted parameter $f_2 = 0.063$ from (\ref{2.18}). Inserting
$W^2 = Q^2/x$, from (\ref{3.8}), we obtain the gluon distribution as
a function of $x$ and $Q^2$.

\begin{figure}[h!]
\begin{center}
\vspace*{-2.3cm}
\includegraphics[scale=.4]{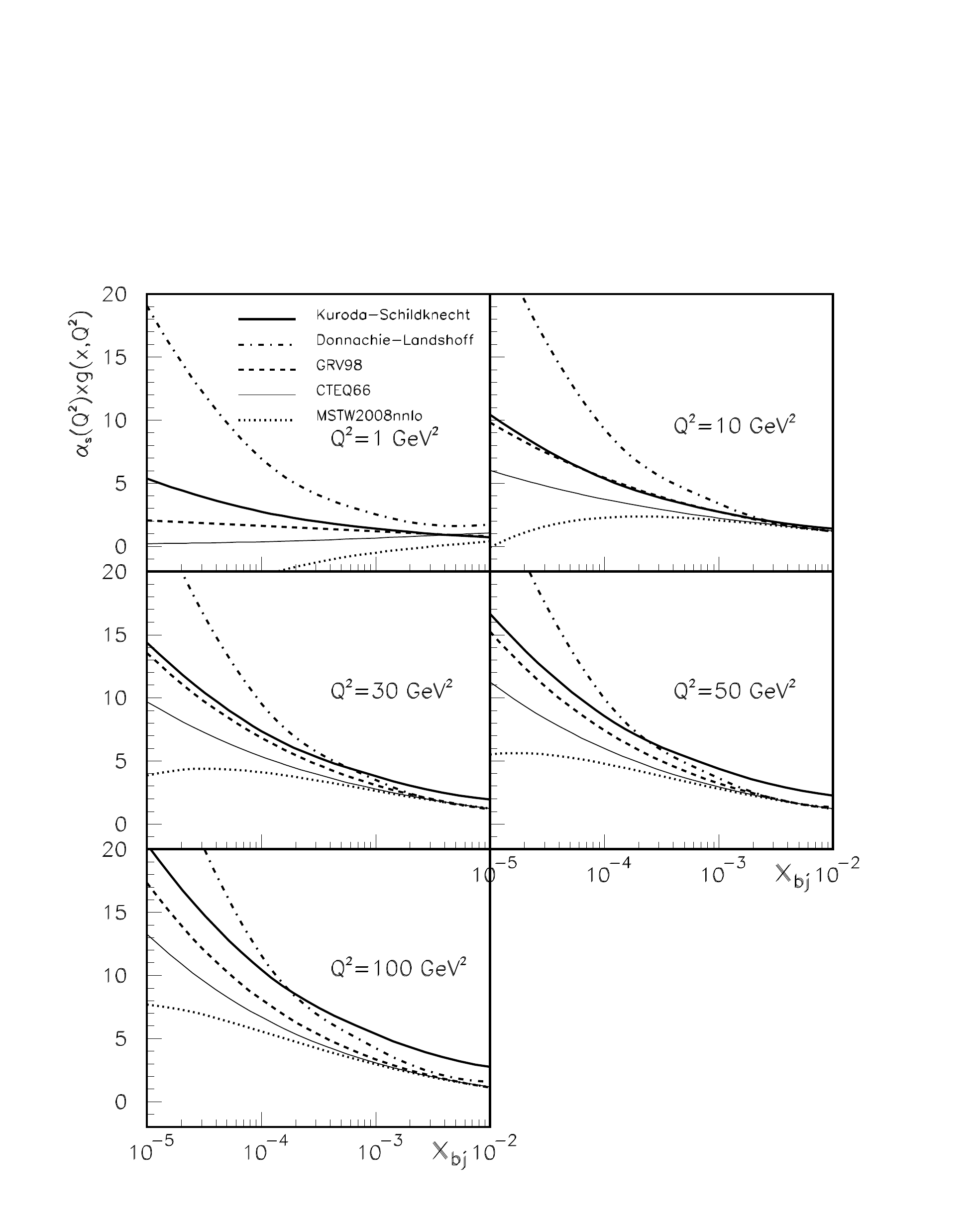}
\vspace*{-0.8cm}
\caption{The gluon distribution function (\ref{3.8}) compared with
the results from the hard Pomeron part of a Regge fit to $F_2 (x,Q^2)$
and from the fits GRV \cite{GRV} CTEQ \cite{CTEQ} and \cite{MSTW}.}
\end{center}
\end{figure}

Using the next-to-leading order expression for $\alpha_s(Q^2)$, in
Fig.\ 11 \cite{E}, we compare the gluon distribution (\ref{3.8}) with
various gluon distributions obtained in sophisticated fits to the 
experimental data. The consistency of our simple one-free-parameter
extraction of the gluon distribution from Fig. 6 according to
(\ref{3.8}) may indicate that the gluon distribution is less sensitively
dependent on the details of the $ggpp$ structure than usually assumed,
or elaborated upon and employed in the global fits to the experimental data.

\section{Specific ansatz for the dipole cross section and comparison with
  experiment}
\renewcommand{\theequation}{\arabic{section}.\arabic{equation}}
\setcounter{equation}{0}

Any specific ansatz for the dipole cross section 
has to interpolate between the
region of $\eta (W^2,Q^2) \gg 1$, where $\sigma_{\gamma^*p} (\eta (W^2,Q^2))
\sim 1/\eta (W^2,Q^2)$, and the region of $\eta (W^2,Q^2) \ll 1$, where
$\sigma_{\gamma^*p} (\eta(W^2,Q^2)) \sim \ln \left( 1/\eta (W^2,Q^2)\right)$,
compare (\ref{2.13}). For the explicit expressions for the ansatz for the
dipole cross section, we refer to refs. \cite{DIFF2000} and \cite{E}.
We only note that the saturation scale, $\Lambda^2_{sat} (W^2) \sim
(W^2)^{C_2}$ in the HERA energy range approximately varies between
$2 GeV^2 \le \Lambda^2_{sat} (W^2) \le 7 GeV^2$, and
restrict ourselves to presenting a comparison with the experimental data.

\begin{figure}[h!]
\begin{center}
\vspace*{-0.1cm}
\includegraphics[scale=.4]{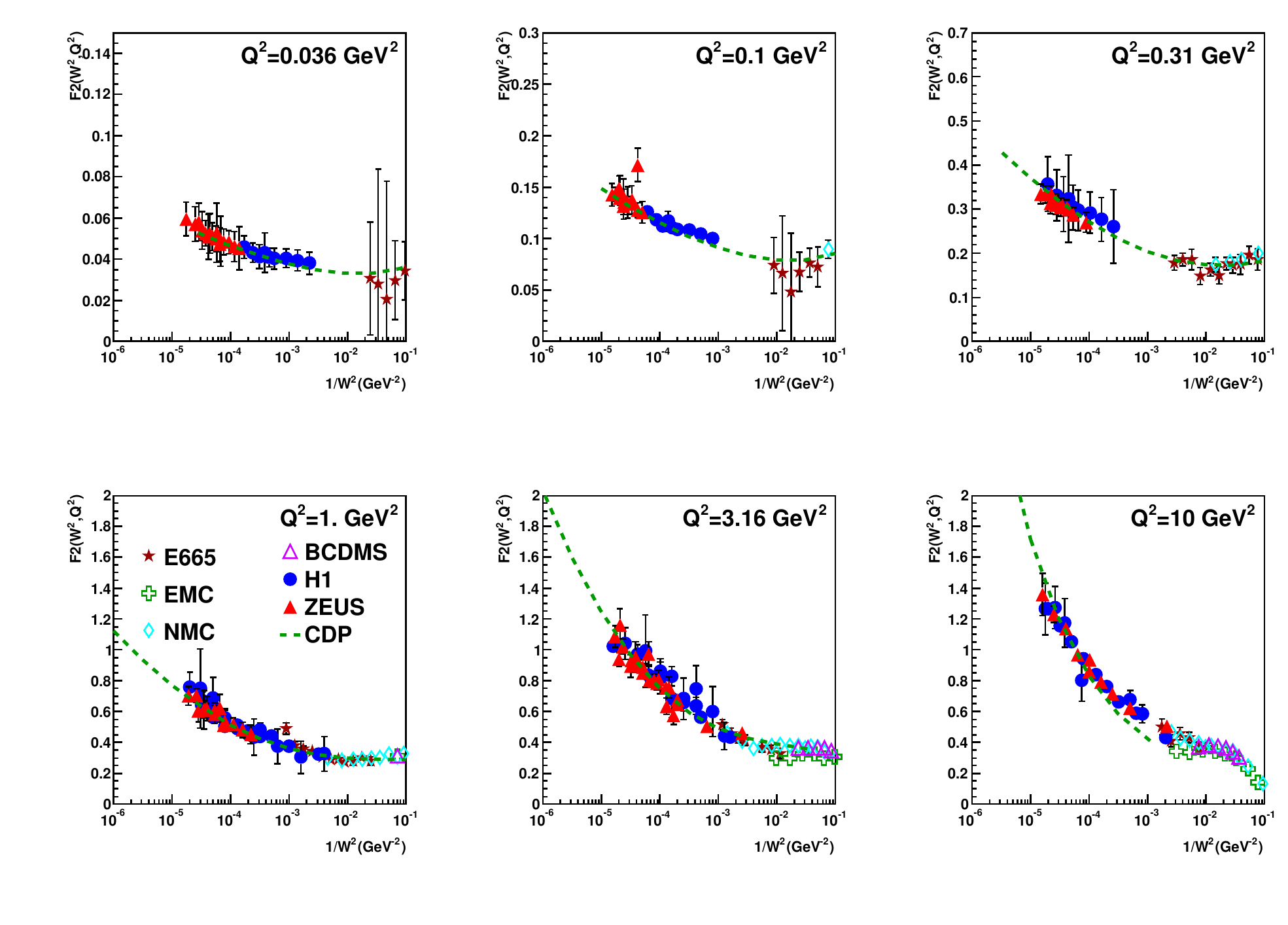}
\vspace*{-1.0cm}
\caption{The predictions \cite{E} from the CDP for the structure function
$F_2 (W^2,Q^2)$ compared with the experimental data for $0.036 \le Q^2 \le
10 GeV^2$.}
\end{center}
\end{figure}
\begin{figure}[h]
\begin{center}
\vspace*{-1.0cm}
\includegraphics[scale=.4]{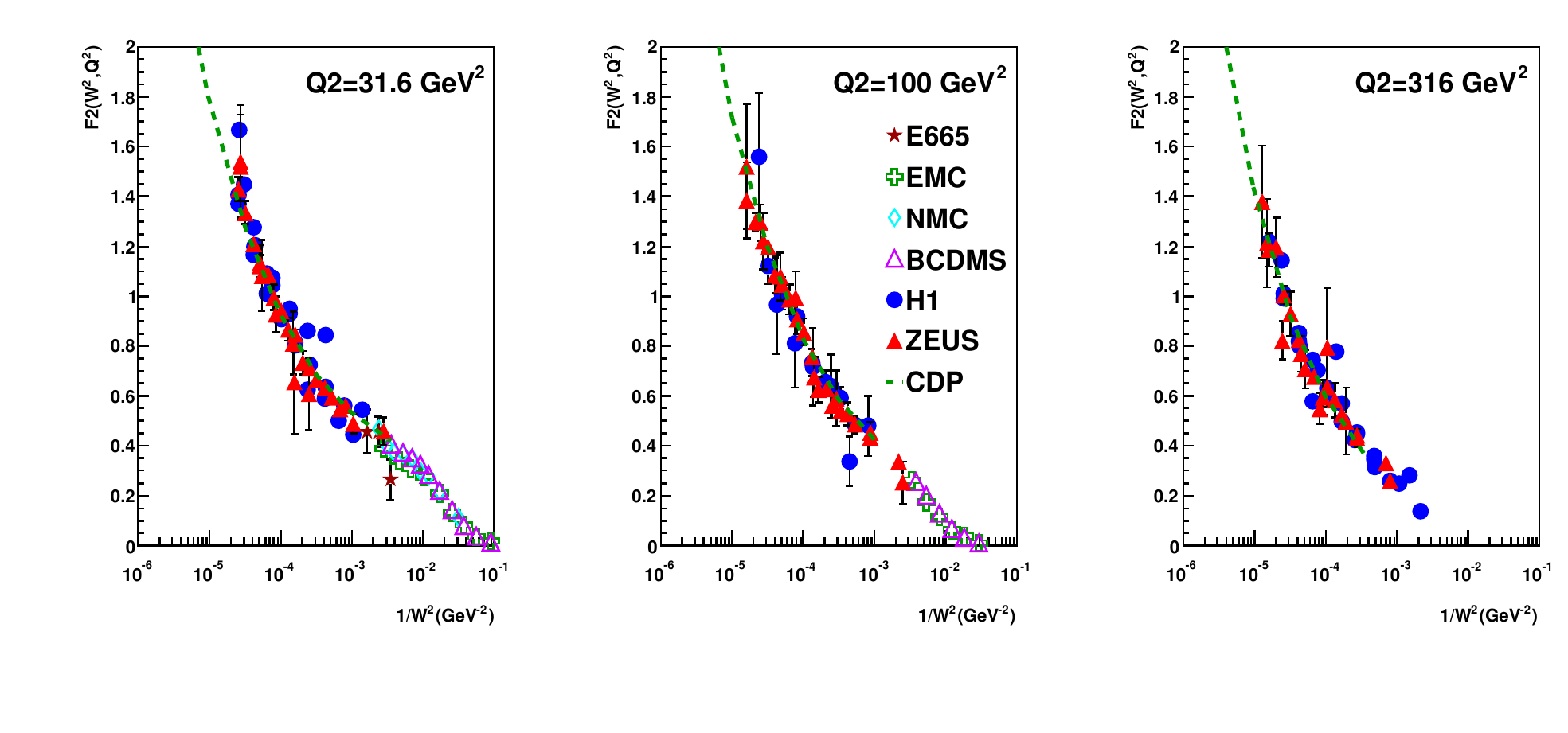}
\vspace*{-1.2cm}
\caption{As in Fig. 12, but for $31.6 GeV^2 \le Q^2 \le 316 GeV^2$.}
\vspace*{-0.2cm}
\end{center}
\end{figure}

The theoretical results from the CDP in Figs. 12 and 13 show agreement
with the experimental data\footnote{We thank Prabhdeep Kaur for providing
the plots of the experimental data in Figs 12 to 15.} 
for $F_2 (W^2,Q^2)$ over the full relevant
region of $0.036~GeV^2 \le Q^2 \le 316 GeV^2$.

In Figs.\ 14 and 15, in addition to the theoretical results from the CDP,
for comparison, we also show the results from the pQCD improved parton
model based on the gluon distribution function (\ref{3.8}) shown in\break
Fig. 11.

Explicitly, by returning to (\ref{2.18}) and inverting (\ref{3.8}), we
reinterprete (\ref{3.8}) as a prediction from the (previously determined)
gluon distribution according to 
\bqa
&& \hspace*{-0.5cm} F_2 (W^2 = Q^2/x) = f_2 
\left( \frac{W^2}{1 GeV^2} \right)^{C_2 = 0.29}
\nonumber \\  
&&\hspace*{-0.5cm} = \frac{(2 \rho + 1) \sum_q Q^2_q}{3 \pi} 
\xi^{C_2=0.29}_L \alpha_s (Q^2) G(x,Q^2),
\label{4.1}
\eqa
where $f_2 = 0.063$ and $\xi_L = 0.40$. In Figs. 14 and 15, we see the 
expected consistency of the
pQCD prediction (\ref{4.1}) with the experimental data 
and the CDP in the 
relevant range of $10 GeV^2 \le Q^2 \le 100 GeV^2$. For $Q^2 < 10 GeV^2$,
with increasing $W$, gradually saturation sets in implying a breakdown of
the pQCD proportionality (\ref{4.1}) of the proton structure function to
the gluon distribution. 

\begin{figure}[h!]
\begin{center}
\includegraphics[scale=.4]{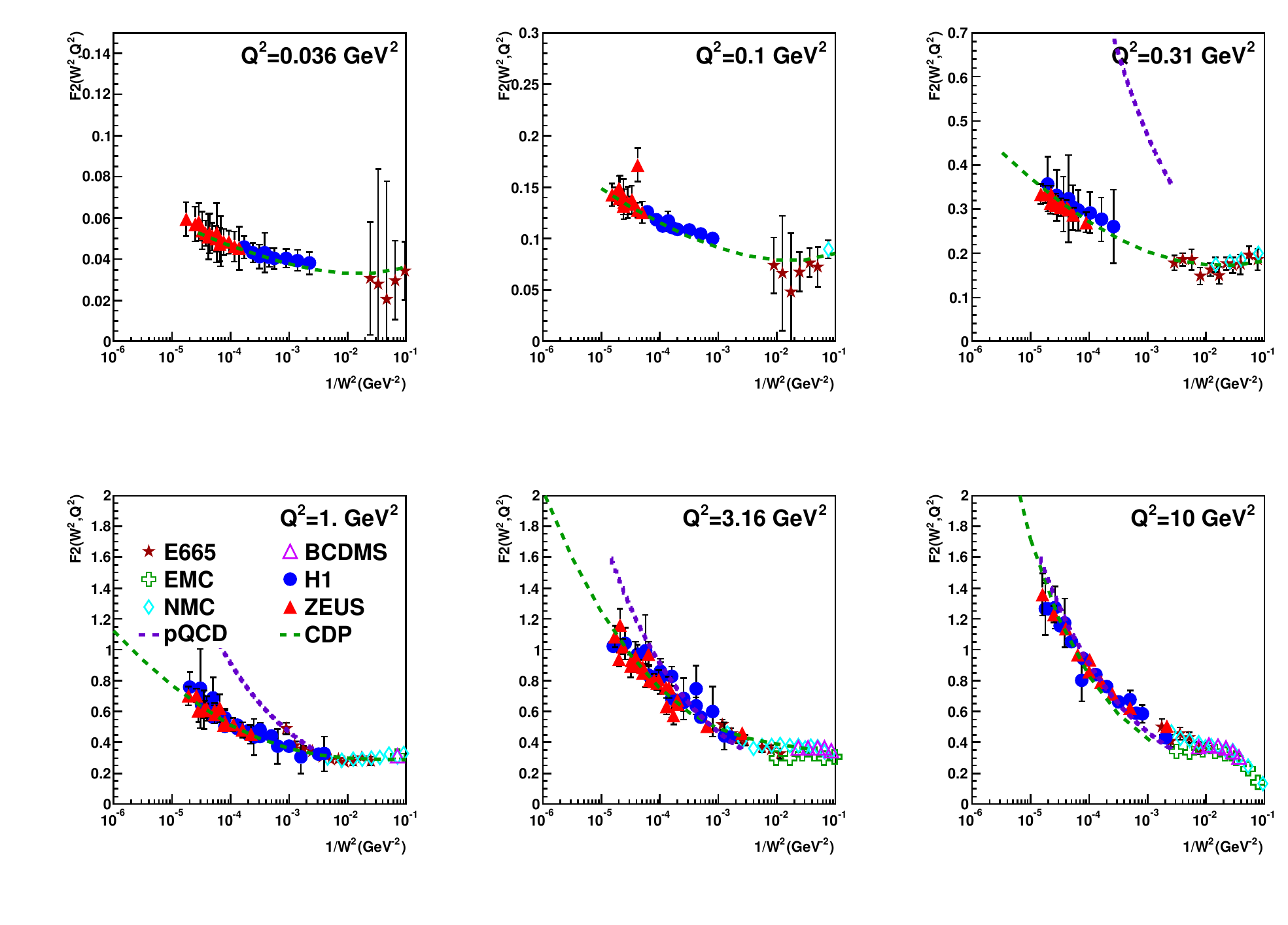}
\vspace*{-1.0cm}
\caption{In addition to the prediction from the CDP, also the pQCD prediction
(\ref{4.1}) based on the gluon distribution (\ref{3.8}) is compared with
the experimental data for $F_2 (W^2,Q^2)$.}
\end{center}
\end{figure}

\begin{figure}[h!]
\begin{center}
\vspace*{-1.0cm}
\includegraphics[scale=.4]{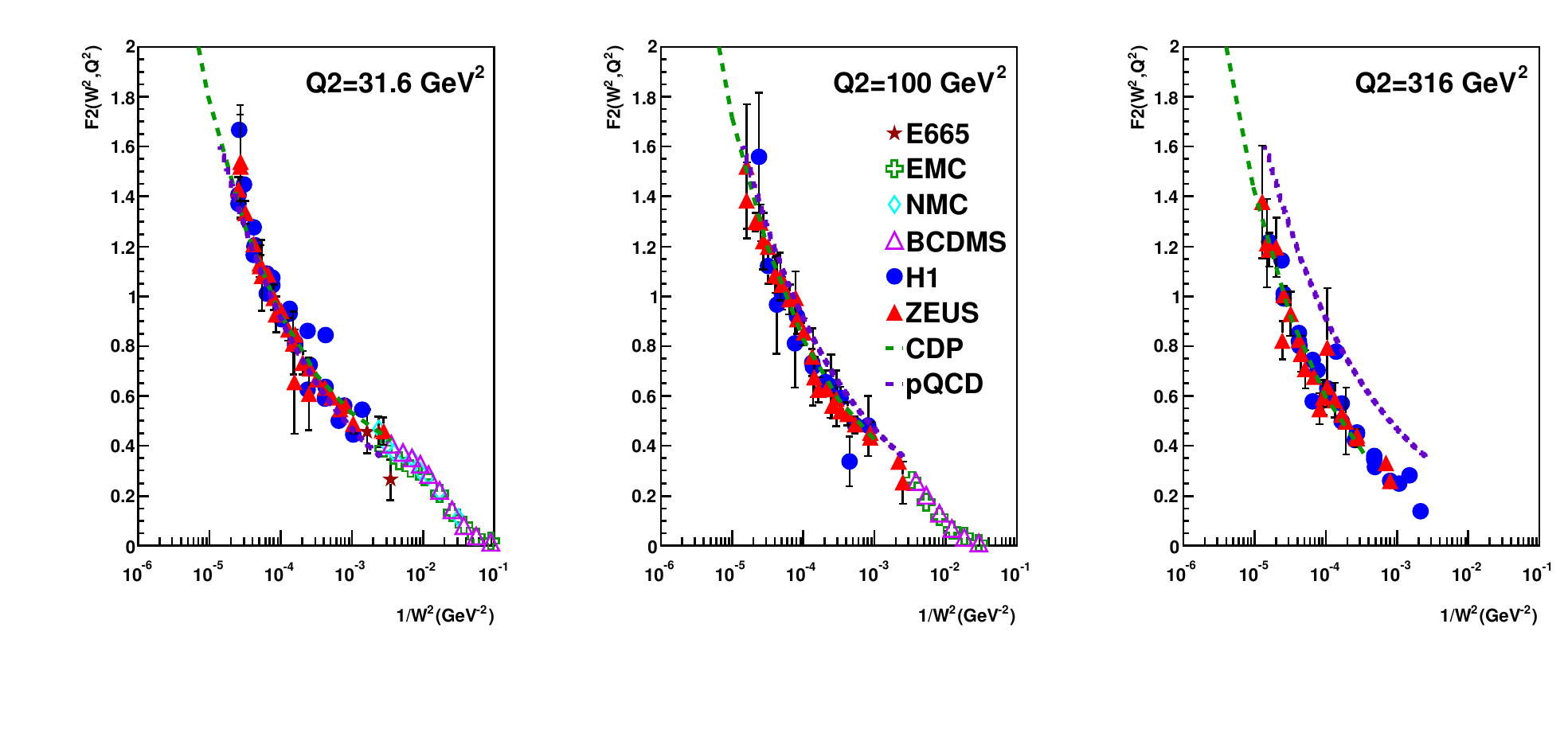}
\vspace*{-1.2cm}
\caption{As in Fig. 14, but for $31.6 GeV^2 \le Q^2 \le 316 GeV^2$.}
\vspace*{-0.4cm}
\end{center}
\end{figure}

The pQCD prediction
for $Q^2 = 316 GeV^2$ lies above the experimental data. This is due to the
breakdown of the simple form for $F_2 (W^2)$ in (\ref{2.18}) which is 
used to extract the gluon distribution. Employing the full CDP result
would lead to an appropriate decrease of the gluon distribution with
increasing $Q^2$ for $Q^2 > 100 GeV^2$.

The proton structure function of the CDP, according to (\ref{2.13}),
in the region of $\eta (W^2,Q^2) 
< 1$, starts to depend logarithmically on the saturation scale,
$\Lambda^2_{sat} (W^2) \sim (W^2)^{C_2}$, and with $\alpha_s (Q^2) G(x,Q^2)
\sim (W^2)^{C_2}$ from (\ref{3.8}) and (\ref{4.1}), it depends
logarithmically on the gluon distribution,
\bqa
&& \hspace*{-1cm} F_2 (W^2,Q^2) \sim Q^2 \sigma_L^{(\infty)} \ln 
\frac{\Lambda^2_{sat}
(W^2 = Q^2/x)}{Q^2 + m^2_0} \nonumber \\
&& \hspace*{-1.2cm} \sim  Q^2 \sigma_L^{(\infty)} \ln \frac{\alpha_s (Q^2) G (x,Q^2)}
{\sigma_L^{(\infty)} (Q^2 + m^2_0)}, \left( Q^2 \ll \Lambda^2_{sat}
  (W^2)\right).
\label{4.2}
\eqa
The smooth transition from the color transparency region to the saturation
region does not correspond to a change in the $W$-dependent gluon
distribution, $\alpha_s (Q^2) G(x,Q^2)$, but occurs via transition from
the proportionality (\ref{4.1}) to the logarithmic  dependence
(\ref{4.2}) on the gluon distribution function. It is the same gluon
distribution that is relevant in the region of color transparency
$\eta (W^2,Q^2) \gg 1$ and in the saturation region, $\eta (W^2, Q^2) \ll 1$,
but the functional
dependence on the gluon distribution has changed. We disagree
with the frequently expressed opinion (compare e.g. ref. \cite{F} and 
the list of references given therein) that the mere
existence of the saturation scale and of scaling like in Fig. 7
suggests and even requires a modification based on non-linear evolution
of the gluon distribution determined in the pQCD domain of 
$\eta (W^2,Q^2) > 1$,
when passing to the saturation region of $\eta (W^2,Q^2) < 1$.

\begin{figure}[h]
\begin{center}
\vspace*{-0.2cm}
\includegraphics[scale=.25]{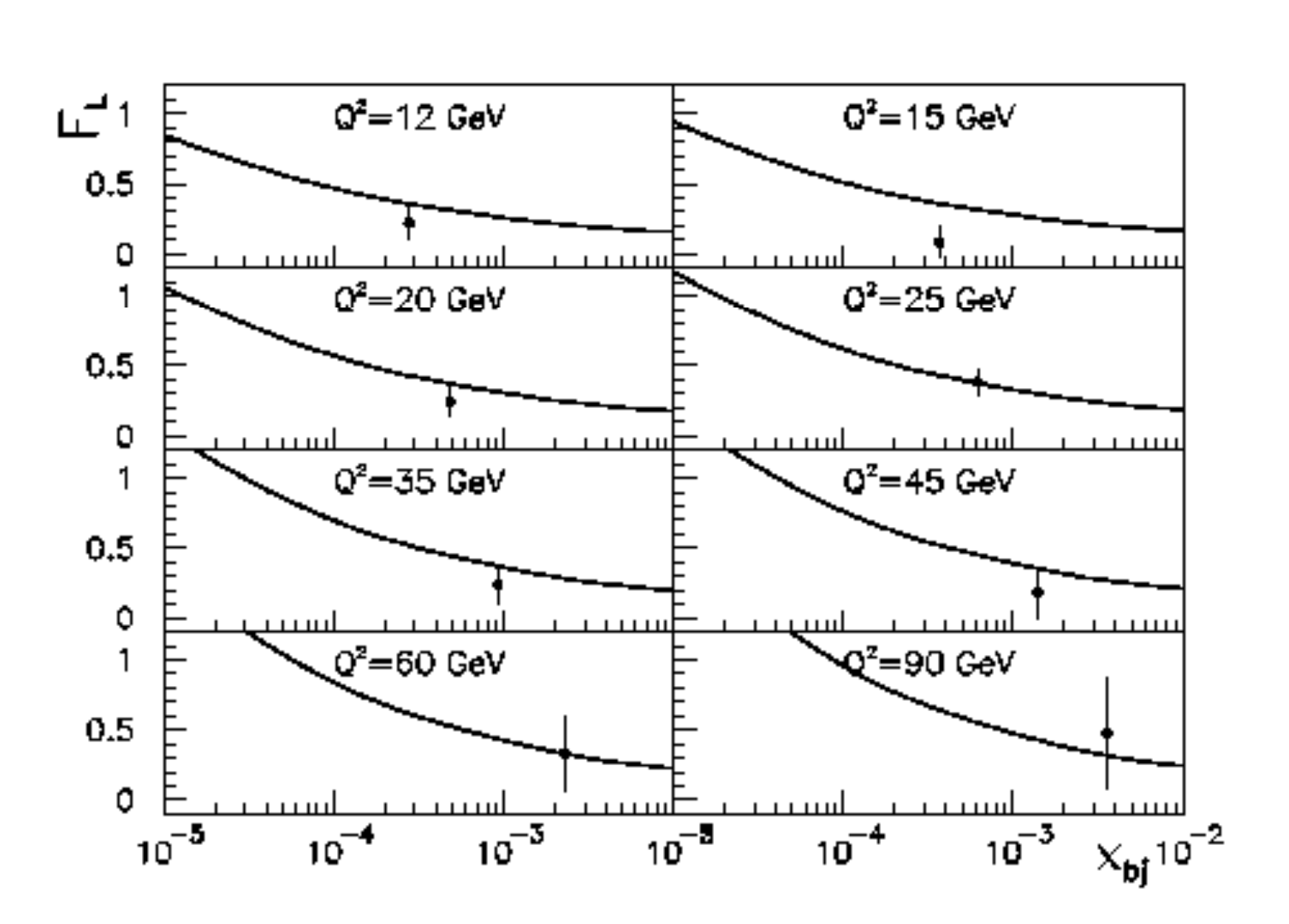}
\includegraphics[scale=.25]{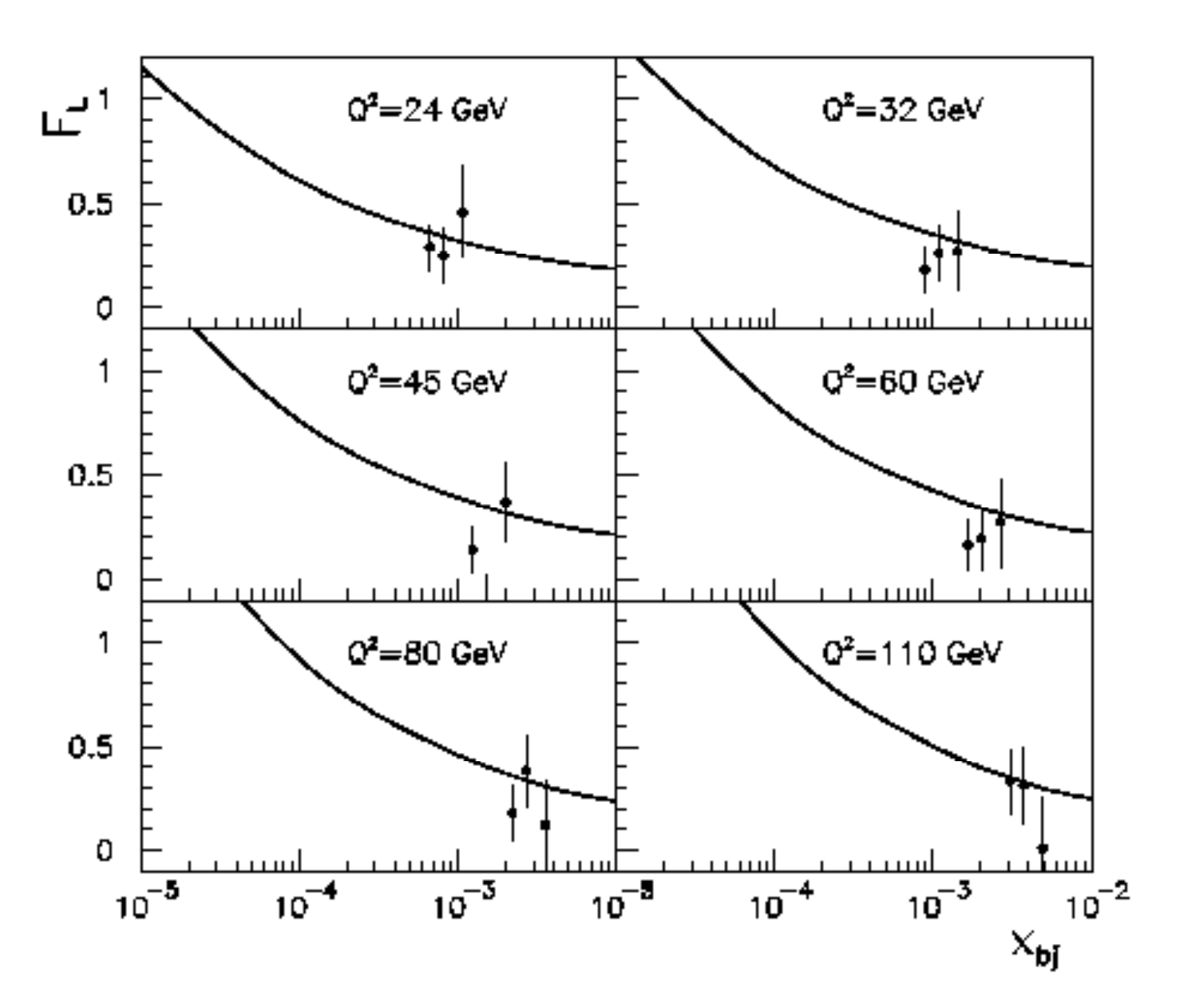}
\vspace*{-0.3cm}
\caption{The longitudinal proton structure function $F_L (x, Q^2)$
from the H1 collaboration \cite{h1data} and from the ZEUS collaboration
\cite{zeusdata} compared with the CDP prediction.}
\vspace*{-0.4cm}
\end{center}
\end{figure}

In Fig. 16, we show the longitudinal structure function, $F_L (x,Q^2)$, in
comparison with the predictions from the CDP based on the specific ansatz
for the dipole cross section that was used in Figs. 12 to 15.





\section{Conclusions}

The color-gauge-invariant dipole interaction with the nucleon 
in terms of the forward scattering amplitude proceeds
via two reaction channels. They correspond to the two diagrams of Fig. 4.
For a given dipole size, if both channels are open, the destructive 
interference between them implies color transparency. In the limit
in which the reaction
channel corresponding to the second diagram in Fig. 4 is closed, the dipole
cross section saturates to a cross section of standard hadronic size with
at most a weak energy dependence.

In the photoabsorption cross section, the above limits of color 
transparency and saturation are realized, respectively, by the 
two regions in the $(Q^2, W^2)$ plane corresponding to $\eta (W^2,Q^2) \gg 1$
and $\eta (W^2, Q^2) \ll 1$. The $(Q^2, W^2)$ plane (under the restriction
of $Q^2/W^2 \lsim ~0.1$) is accordingly simple. There are only two distinct
regions, separated by the line $\eta (W^2,Q^2) = 1$.

The main features of the experimental data on DIS at low x have thus been
recognized to follow from the color-gauge-invariant dipole-proton
interaction without adopting a specific ansatz for the dipole cross section.
Any specific ansatz has to interpolate between the model-independent
restrictions on the photoabsorption cross section that hold for
for $\eta (W^2, Q^2) \gg 1$ and for
$\eta (W^2,Q^2) \ll 1$, respectively.

\section*{Acknowledgement}

The author is grateful to Kuroda-san for a fruitful collaboration on the
color dipole picture. Thanks to Prabhdeep Kaur for providing plots of
experimental data. The author thanks Allen Caldwell and Reinhart
K\"ogerler for useful discussions.

The lively scientific and pleasant atmosphere at Ring Castle is gratefully
acknowledged.



\begin{thebibliography}{00}

\bibitem{A}
  D. Schildknecht, Acta Phys. Polon. {\bf B37} (2006) 595
  [hep-ph/0511090].

\bibitem{B}
  L. Stodolsky, Phys. Rev. Lett. {\bf 18} (1967) 135.  

\bibitem{Sakurai} 
  J.J. Sakurai and D. Schildknecht, Phys. Lett. {\bf 40B}
  (1972) 121;\hfill\break
  B. Gorczyca, D. Schildknecht, Phys. Lett. 47B, 71 (1973).

\bibitem{Fraas}
  H. Fraas, B.J. Read and D. Schildknecht, Nucl. Phys. {\bf B86}
  (1975) 346;\hfill\break
  R. Devenish and D. Schildknecht, Phys. Rev. {\bf D19} (1976) 93.

\bibitem{C} 
  EMC-NMC Collaboration, compare: M. Arneodo, Phys. Rep.{\bf 240}
  (1994) 301.

\bibitem{D}
  D. Schildknecht, Nucl. Phys. {\bf B66} (1973) 398;\hfill\break
  C. Bilchak and D. Schildknecht, Phys. Lett. {\bf B214} (1988)
  441; Phys. Lett. {\bf B233} (1989) 461.

\bibitem{Low} 
F.E. Low, Phys. Rev. {\bf D12} (1075) 163;\hfill\break
S. Nussinov, Phys. Rev. Lett. {\bf 34} (1975)  1286; 
Phys. Rev. {\bf D14} (1976) 246;
\hfill\break
J. Gunion, D. Soper, Phys. Rev. {\bf D15} (1977) 2617.

\bibitem{Nikolaev} 
N.N. Nikolaev, B.G. Zakharov, Z. Phys. C49, 607 (1991). 

\bibitem{Cvetic} 
G. Cvetic, D. Schildknecht, A. Shoshi, Eur. Phys. J {\bf C13}
(2000) 301.

\bibitem{DIFF2000} 
D. Schildknecht, Contribution to Diffraction 2000, Cetraro,
Italy, September 2-7, 2000, Nucl. Phys. B, Proc. Supplement {\bf 99} (2001)
121;
\hfill\break
D. Schildknecht, B. Surrow, M. Tentyukov, Phys. Lett. {\bf B499} (2001) 116;
\hfill\break
G. Cvetic, D. Schildknecht, B. Surrow, M. Tentyukov, EPJC {\bf 20} (2001) 77.

\bibitem{Forshaw}
J.R. Forshaw, G. Kerley, G. Shaw, Phys. Rev. {\bf D60} (1999) 074012.

\bibitem{Ewerz} 
C. Ewerz and O. Nachtmann, Annals of Phys. {\bf 322} (2007) 1635; 
{\bf 322} (2007) 1670; \\
C. Ewerz, A.v. Manteuffel, O. Nachtmann, [arXiv:1101.028 [hep-ph]].

\bibitem{MKDS} M. Kuroda and D. Schildknecht, Phys. Rev. {\bf D66}(2002) 094005;
Phys. Rev. {\bf D67} (2003) 094008.   


\bibitem{E} 
M. Kuroda and D. Schildknecht, [arXiv: 1108.2584 [hep-ph]]

\bibitem{Ku-Schi} 
M. Kuroda and D. Schildknecht, Phys. Lett. {\bf B670} (2008) 129.  

\bibitem{ST} 
D. Schildknecht and F. Steiner, Phys. Lett. {\bf B56} (1975) 36.

\bibitem{SCHI}
D. Schildknecht, Contribution to DIS 2001, The 9th International
Workshop on Deep Inelastic Scattering, Bologna, Italy, 2001, G. Brassi et al.
(Eds.), World Scientific, Singapore, 2002, p. 798;\hfill\break
D. Schildknecht, B. Surrow and M. Tentyukov, Mod. Phys. Lett. {\bf A16} 
(2001) 1829.

\bibitem{CAL} 
A. Caldwell, [arXiv:0802.0769]

\bibitem{23a}
D. Schildknecht, [arXiv:1104.0850 [hep-ph]]

\bibitem{Martin}
A.D. Martin, R.G. Roberts and W.J. Stirling, Phys. Rev. {\bf D37} (1988) 1161; \\
A.M. Cooper-Sarkar et al., Z. Phys. {\bf C39} (1988) 281.

\bibitem{Lipatov}
L.N. Lipatov, Sov. J. Nucl. Phys. {\bf 20} (1975) 95, \\
V.N. Gribov and L.N. Lipatov, Sov. J. Nucl. Phys. {\bf 15} (1972) 438, \\
G. Altarelli and G. Parisi, Nucl. Phys. {\bf B126} (1977) 298; \\
Yu. L. Dokshitzer, Sov. Phys. JETP {\bf 46} (1977) 641.

\bibitem{Prytz}
K. Prytz, Phys. Lett. {\bf B311} (1993) 286.

\bibitem{Adel}
K. Adel, F. Barreiro, F.J. Yndurain, Nucl. Phys. {\bf B495} (1997) 221; \\
F.J. Yndurain, The Theory of Quark and Gluon Interactions (Springer 1999), p.157.

\bibitem{Dom}
A. Donnachie and P.V. Landshoff, Phys. Lett. {\bf B533} (2002) 277,
[hep-ph/0111427]; 
Acta Physica Polonica {\bf B34} (2003) 2989, [hep-ph/0305171].

\bibitem{GRV}
M. Gl\"uck, E. Reya, A. Vogt, 
Z. Phys. C67 (1994) 433; 
Eur. Phys. J. C5 (1998) 461

\bibitem{CTEQ}
CTEQ Collaboration: J. Pumplin et al., JHEP 0207 (2002) 012.

\bibitem{MSTW}
A.D. Martin et al., Eur. Phys. J. C18 (2000) 117.

\bibitem{F}
J. Kuokkanen, K. Rummukainen and H. Weigert,
[arXiv:1108.1867 [hep-ph]].

\bibitem{h1data}
H1 Collaboration, F.D. Aaron et al., Phys. Lett. B665 (2008) 139.

\bibitem{zeusdata}
ZEUS Collaboration, S. Chekanov et al., Phys. Lett. B682(2009) 8.

\end{thebibliography}




\end{document}